\documentclass[11pt, a4paper]{article}

\pdfoutput=1
\usepackage{bbold}
\usepackage{jheppub}
\usepackage{times}
\usepackage{mathrsfs} 
\usepackage[english]{babel}
\renewcommand{\vec}[1]{\mathbf{#1}} 


\title{Accounting for the Heisenberg and Pauli principles in the kinetic approach to neutrino oscillations}

\author{A. Kartavtsev\\}

\affiliation{P. G. Demidov Yaroslavl State University, Sovietskaya 14, 150003 Yaroslavl, Russia}

\emailAdd{alexander.kartavtsev@mpp.mpg.de}

\arxivnumber{2007.13736}

\makeatletter
\gdef\@fpheader{JHEP11(2020)135}
\makeatother
 
\abstract{While oscillations of solar neutrinos are usually studied using the single-particle quan\-tum-mechanical approach, flavor conversions of supernovae neutrinos are typically analyzed using the kinetic equation for the matrix of densities due to the necessity of including also the scattering processes. Using the Wigner formulation of quantum mechanics we show the equivalence of the quantum-mechanical and kinetic approaches in the limit of collisionless neutrino propagation (in a background medium). Based on this observation we also argue that solutions of the kinetic equation account for the Heisenberg uncertainty principle and the related effect of wave packet separation (for single neutrinos), as well as the Pauli exclusion principle, if the initial conditions are consistent with these fundamental quantum principles. Such initial conditions can be constructed e.g. by iden\-tifying the matrix of densities with the (reduced) single-particle Wigner function computed using initial conditions for the neutrino wave function. Hence the neutrino momentum un\-certainty is an integral part of the initial conditions for the matrix of densities, that  may have an impact on the phenomenology of supernovae neutrinos via the effect of wave packet separation.}

\begin{document}

\maketitle
\flushbottom
\newpage

\section{\label{sec:introduction}Introduction}
Neutrinos play an important role in the stellar dynamics and provide us with an invaluable tool to study stellar evolution. The detection of neutrinos produced by the Sun, pioneered by Ray Davis and his collaborators \citep{Davis:1968cp} and later verified by the Super-Kamiokande \citep{Fukuda:2002pe} and SNO \cite{Ahmad:2002jz} experiments, confirmed the Standard Solar Model \citep{Bahcall:2000nu} and led to the discovery of flavor neutrino oscillations. The observation of neutrinos from the supernova SN1987A by the Kamiokande II \citep{Hirata:1987hu}, IMB \citep{Bionta:1987qt}, and BNO \citep{Alekseev:1987ej} collaborations provided an insight into the process of the core collapse and the sub\-se\-quent formation of a neutron star, as well as put constrains on properties of Beyond the Standard Model physics. The long-awaited \citep{Antonioli:2004zb} detection of new bursts is expected to provide further valuable infor\-mation on the mechanism of supernovae explosion.

Neutrino propagation in the solar interior is almost collisionless and can be described by the Schr\"odinger equation for the neutrino wave function. Although scattering processes changing the neutrino momentum are practically irrelevant,  coherent forward scattering on the ambient matter strongly affects flavor conversion of solar neutrinos through the MSW effect \citep{Wolfenstein:1977ue,Mikheev:1986gs}. Furthermore, the effect of decoherence by wave packet separation, related to the neutrino momentum uncertainty and thus to the Heisenberg uncertainty principle, that is naturally accounted for in the quantum-mechanical approach, damps the oscillations. As a result, even a hypothetical experiment detecting one neutrino at a time would not observe a seasonal variation of the neutrino flavor composition. 

On the other hand, in some phases of supernovae evolution neutrino collisions with particles of the ambient medium can play a dominant role. Due to the necessity of including also the scattering processes, flavor conversions of supernovae neutrinos are usually analyzed using the kinetic equation for the matrix of densities \citep{Sigl:1992fn}. The oscillation term of the kinetic equation naturally incorporates coherent forward scattering and the related MSW effect. The Pauli-blocking factors in its collision term ensure that the exclusion principle is respected in the scattering processes. However, because the matrix of densities is a function of coordinate and momentum, the kinetic equation is often perceived as classical, i.e. describing an ensemble of particles with definite coordinates and momenta. For this reason, its consistency with the  uncertainty principle and the ability to account for the effect of wave packet separation (at the level of individual neutrinos) are questioned. 
 
The goals of the present work are twofold. 
First, we show that for collisionless neutrino propagation (in a background medium) the quantum-mechanical and kinetic approaches to neutrino oscillations produce equivalent results. Second, based on this observation, 
we argue that solutions of the kinetic equation account for the Heisenberg  principle and the related effect of wave packet se\-p\-aration, as well as the Pauli principle, if the initial conditions are consistent with these fundamental quantum principles. To this end, we make use of the Wigner formulation of quantum mechanics and show that the form of the evolution equation for the (reduced) single-particle Wigner function matches the form of the kinetic equation for the matrix of densities. Constructing initial conditions for the matrix of densities using the neutrino wave function we also show that their form matches the form of the initial conditions for the (reduced) single-particle Wigner function. Therefore, (in the collisionless limit) the quantum-mechanical and kinetic approaches to neutrino oscillations produce equivalent results. Because the quantum-mechanical approach accounts for the uncertainty and exclusion principles, the same applies also to the kinetic approach provided that the kinetic equation is supplemented by appropriate initial conditions.

This line of reasoning is elaborated on in sections \ref{sec:qmapproach} to \ref{sec:kinapproach}.  In section \ref{sec:qmapproach} we `translate' the by now standard quantum-mechanical approach to solar neutrinos into the language of the single-particle Wigner function. Furthermore, we derive an evolution equation for the Wigner function and show that it accounts for the uncertainty principle if supplemented by initial conditions constructed from the neutrino wave function. In section \ref{sec:manypartqm} we generalize this analysis to $N$-particle neutrino systems. In particular, we derive an evolution equation for the reduced single-particle Wigner function and show that it accounts for the exclusion principle if supplemented by initial conditions constructed from the $N$-particle wave function. In section \ref{sec:kinapproach} we derive the kinetic equation in the limit of col\-li\-sionless neutrino propagation (but including coherent forward scattering) and show that its form matches the form of the evolution equation for the (reduced) single-particle Wigner function. We also construct initial conditions for the matrix of densities using the neutrino wave function and show that they match those for the Wigner function. The obtained results are summarized in section \ref{sec:conclusions}. Finally, appendices \ref{sec:apprthreeflavor} to \ref{sec:densitymatrix} contain supplemen\-tary technical material. 

\section{\label{sec:qmapproach}Single-particle quantum-mechanical approach}
Thermonuclear reactions that power the Sun produce a large flux of electron neutrinos. Due to the flavor conversion that occurs during neutrino propagation in the Sun, in vacuum, and in the Earth neutrino flavor composition measured by terrestrial experiments differs from that at the production point. As the neutrino propagation in these three environments is (almost) collisionless, its evolution is well described by the Schrödinger equation. The Wigner function constructed from solutions of the latter can be represented as a convolution of a shape and a phase factor. The shape factor describes position of the neutrino wave packet and satisfies the Liouville equation. The phase factor describes con\-version of the neutrino flavor and satisfies a Schrödinger-like equation. A rather accurate estimate of the phase factor can be obtained taking into account that for the experimentally known neutrino parameters:
	i) mixing angle $\theta_{13}$ and the medium corrections to $\theta_{13}$ are small;
	ii) neutrino propagation in the Sun is close to adiabatic;
	iii) neutrino propagation in the Earth is close to adiabatic within the layers, but the adiabaticity is maximally violated at the boundaries between the layers. Due to the effect of decoherence by wave packet separation solar neutrinos arrive at the Earth as an incoherent superposition of the mass eigenstates. For this reason their flavor composition does not exhibit a seasonal variation. Crossing the Earth surface, the neutrino mass eigenstates split into the matter propagation eigenstates  and start to oscillate. This leads to a small difference between the flavor composition of the night-time and day-time neutrinos.  Below we briefly review the single-particle  quantum mechanical approach to solar neutrinos paying particular attention to the interplay of the shape and phase factors and the averaging prescription, used to take into account details of the neutrino energy spectrum and of the spatial distribution of the production points.

\begin{figure}[t!]
\includegraphics[width=0.49\textwidth]{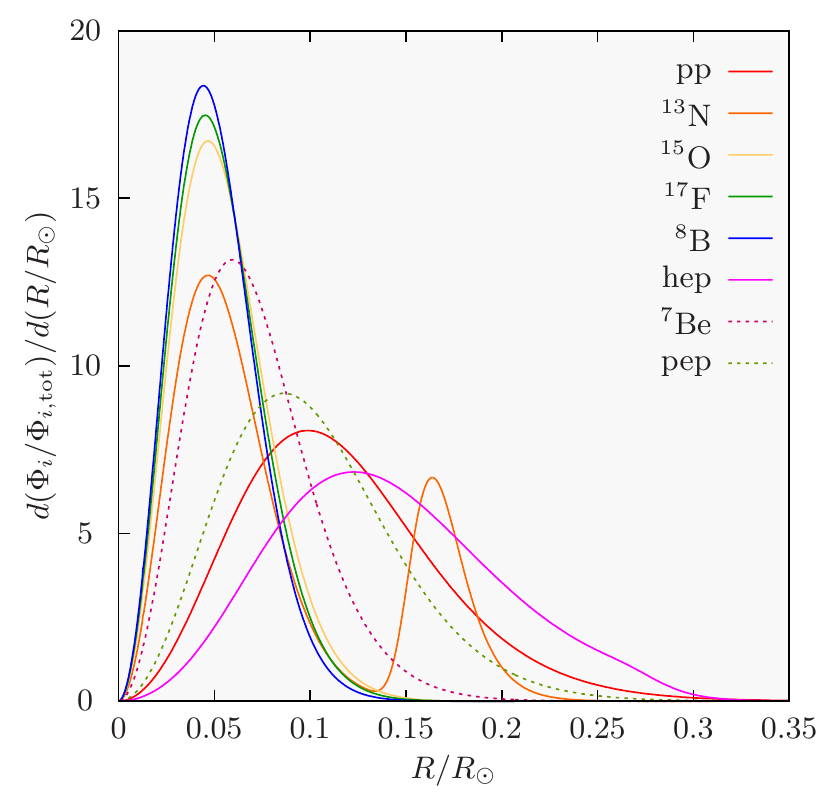}
\includegraphics[width=0.495\textwidth]{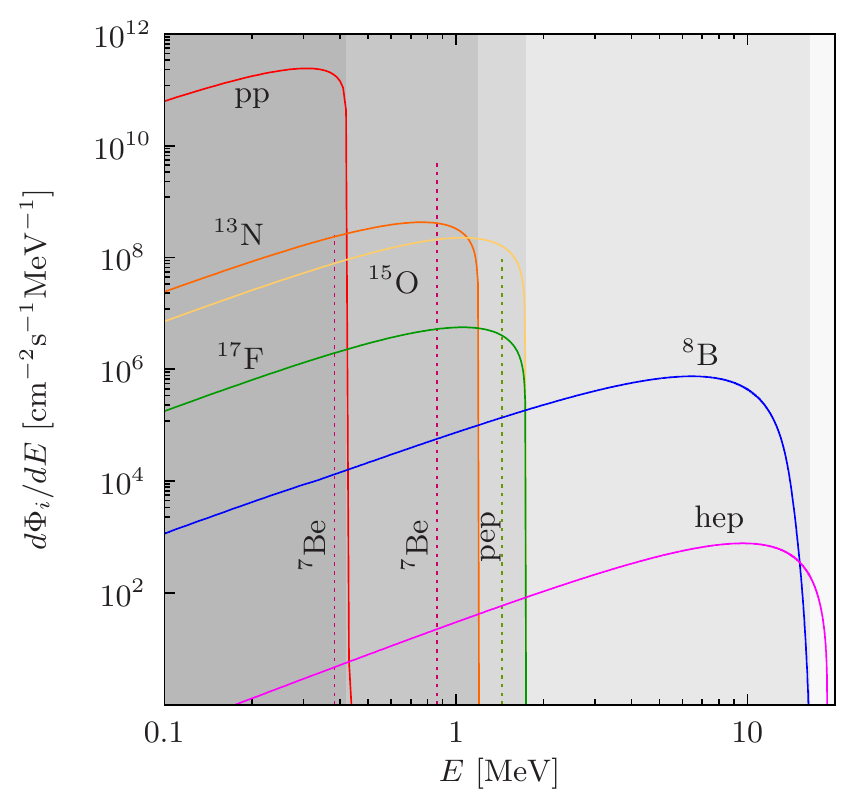}
\caption{\label{fig:spectrum} \emph{Left panel:} Predictions of the Standard Solar Model BPS08 \citep{PenaGaray:2008qe} for the  radial distribution of the neu\-trino fluxes produced in the $pp$-chain and CNO cycle. \emph{Right panel:} Energy spectra of the individual flux components. The gray-shaded areas mark energy intervals dominated by a particular flux component.}
\end{figure}

\paragraph{Solar neutrino fluxes.}
The Sun is a main-sequence star. Roughly 99\% of its energy is produced in the so-called $pp$-chain, while the remaining 1\% is produced in the CNO cycle \citep{PhysRev.55.434}. Both chains transform protons into helium simultaneously creating electron neutrinos, see e.g. figure 10 of reference \citep{Giunti:2007ry}. The neutrino fluxes produced by the individual branches of the $pp$-chain and CNO cycle (the latter recently observed by the Borexino collaboration \citep{Agostini:2020mfq}) are predicted by the standard solar model \citep{Bahcall:2000nu}. Their radial distribution and energy spectra, calculated in reference \citep{PenaGaray:2008qe}, are shown in figure \ref{fig:spectrum}. 

In addition to the energy spectrum characterizing the entire neutrino ensemble, individual neu\-trinos are characterized by the size of their wave packets. The latter is determined by the production process  and differs for different flux components. For the $pp$ and $^7$Be neutrinos $\sigma_x \sim 6\cdot 10^{-8}$~cm \citep{deHolanda:2003nj}. On the other hand, for the $^8$B neutrinos the estimated size of the wave packet is a factor of two larger, $\sigma_x \sim (1-2)\cdot 10^{-7}$~cm \citep{deHolanda:2003nj}. This corresponds to the energy scale $\sigma_p\sim 100$~eV. 

\paragraph{Formal solution of the Schr\"odinger equation.}
Because propagation of solar neutrinos is almost collisionless, its is well described by the Schrödinger equation. In the neutrino mass basis it reads
\begin{align}
\label{eq:schrodeq}
i\partial_t\psi_i(t,\vec{x}) = {\sf H}_{ij}(t,\vec{x})\psi_j(t,\vec{x})\,,
\end{align}
with ${\sf H}_{ij}(t,\vec{x}) \!=\! {\sf K}_{ij}(\vec{x})+{\sf V}_{ij}(t,\vec{x})$ and the kinetic energy operator given by 
${\sf K}_{ij}=\delta_{ij}(-\partial^2_{\vec{x}}+m_i^2)^\frac12$.

Because the width of the neutrino wave packet in coordinate space is usually much smaller than the length scale for the variation of the matter potential in the interior of the Sun or the Earth, one can approximate neutrino propagation by propagation in a spatially homogeneous but time-dependent potential $\mathsf{V}_{ij}(t)$. The value of this effective matter potential at a time $t$ is determined by the potential at the position the neutrino wave packet reaches by that time. This approximation is widely used in the literature. Let us emphasize that it is used in the present section to analyze solar neutrinos, but is not relied upon in sections \ref{sec:manypartqm} and \ref{sec:kinapproach}. For a homogeneous but time-dependent potential the momentum-representation counterpart of the Sch\"rodinger equation \eqref{eq:schrodeq} reads
\begin{align}
\label{eq:schrodmomentum}
i\partial_t\psi_i(t,\vec{p}) = {\sf H}_{ij}(t,\vec{p})\psi_j(t,\vec{p})\,,
\end{align}
where ${\sf H}_{ij}(t,\vec{p}) = \delta_{ij}\,\omega_i(\vec{p}) +  {\sf V}_{ij}(t)$. As can be verified by substitution, its (formal) solution reads
\begin{align}
\label{eq:psiformalsol}
\psi(t,\vec{p})={\sf U}(t,t_P,\vec{p})  \psi(t_P,\vec{p})\,,
\end{align}
where $t_P$ is the neutrino production time and the time evolution operator is given by
\begin{align}
\label{eq:evolutionoperator}
{\sf U}(t,t_P,\vec{p}) = T_c\, \exp\left(-i\int^t_{t_P} {\sf H}(\tau,\,\vec{p})d\tau\right)\,.
\end{align}

The Sun produces only electron-flavor neutrinos. The neutrino flavor states are linear combinations of the mass eigenstates, $\psi_\alpha = U_{\alpha i}\psi_i$, where $U$ is the Pontecorvo-Maki-Nakagawa-Sakata (PMNS) matrix \citep{Pontecorvo:1957qd,Maki:1962mu}. Expressions for wave packets of the neutrino mass eigenstates have been derived in reference \citep{Akhmedov:2010ms} by analyzing the neutrino production in the QFT framework and assuming that the external particles are described by Gaussian wave packets. In particular, it has been found that the (effective) momentum uncertainties of the produced neutrino state are different in different directions. However, because the resulting flavor composition of the day-time and night-time neutrinos does not depend on the exact shape of the neutrino wave packets, here we follow reference \citep{Hansen:2016klk} and neglect the difference of the momentum uncertainties in different directions as well as the difference in the shape of $\psi_i(t_P,\vec{p})$ for different mass eigenstates. In this approximation the initial wave function is given in the flavor basis by
\begin{align}
\label{eq:gaussianwp}
\psi_\alpha(t_P,\vec{p}) = \delta_{\alpha e}\cdot \left(\frac{2\pi}{\sigma^2_p}\right)^\frac34 \exp\left(-\frac{(\vec{p}-\vec{p}_w)^2}{4\sigma_p^2}\right)\,e^{-i\vec{p}\vec{x}_P}\,,
\end{align} 
where $\sigma_p$ is the effective momentum uncertainty of the produced neutrino and $\vec{p}_w$ -- its characteristic momentum. As we will see below, at $t=t_P$ the wave packet is centered at $\vec{x}=\vec{x}_P$.

\paragraph{Single-particle Wigner function.}
The collision integral, that describes neutrino interactions at a particular point $\vec{x}$ of the detector, contains an integration over the neutrino momentum $\vec{p}$. To treat $\vec{x}$ and $\vec{p}$ on equal footing it is convenient to describe neutrino propagation and detection in terms of the Wigner function \citep{Wigner:1932eb}. A single-particle Wigner function, written in terms of the momentum-representation wave function, reads
\begin{align}
\label{eq:wignerfunctionmom}
\varrho_{ij}(t,\vec{x},\vec{p}) = \int \frac{d^3\vec{\Delta}}{(2\pi)^3} e^{i\vec{\Delta}\vec{x}} 
\psi_i(t,\vec{p}+\vec{\Delta}/2)\psi^{*}_j(t,\vec{p}-\vec{\Delta}/2)\,.
\end{align}
Substituting the formal solution equation \eqref{eq:psiformalsol} into equation \eqref{eq:wignerfunctionmom} we arrive at
\begin{align}
\label{eq:wignerfunctimedeptpot}
\varrho(t,\vec{x},\vec{p})&=\!\int\!\frac{d^3\vec{\Delta}}{(2\pi)^3}e^{i\vec{\Delta}\vec{x}}\,
{\sf U}(t,t_P,\vec{p}+{\textstyle \frac12}\vec{\Delta})\psi(t_P,\vec{p}+{\textstyle \frac12}\vec{\Delta}) \nonumber \\
& \times \psi^\dagger(t_P,\vec{p}-{\textstyle\frac12}\vec{\Delta}) {\sf U}^\dagger(t,t_P,\vec{p}-{\textstyle \frac12}\vec{\Delta})\,.
\end{align} 
Note that $\varrho$, $\mathsf{U}$, $\mathsf{H}$ and $\psi$ without generation indices denote matrices and `vector' in the generation space.
To leading order in the gradients ${\sf H}(t,\vec{p}\pm\vec{\Delta}/2) \approx {\sf H}(t,\vec{p}) \pm \partial_{\vec{p}}{\sf H}(t,\vec{p})\vec{\Delta}/2$ where, given the form of the Hamiltonian, $\partial_{\vec{p}}{\sf H}(t,\vec{p}) = \partial_\vec{p}\omega(\vec{p}) \equiv \vec{v}(\vec{p})$. Because solar neutrinos are ultrarelativistic, $\vec{v}(\vec{p})$ can be approximated by the unit vector in the direction of the neutrino propagation, which we denote by $\vec{v}_\vec{p}$ in the following, the corrections being of the order of  $m^2/\vec{p}^2$. In this approximation the velocity matrix is proportional to the identity matrix and hence commutes with the Hamiltonian. From equation \eqref{eq:evolutionoperator} it then follows that ${\sf U}(t,t_P,\vec{p}\pm{\textstyle \frac12}\vec{\Delta})\approx {\sf U}(t,t_P,\vec{p})e^{\mp \frac{i}{2} \vec{\Delta}\vec{v}_\vec{p}(t-t_P)}$. The resulting approximate Wigner function factorizes into a product a phase and a shape factors \citep{Kartavtsev:2014mea},
\begin{align}
\label{eq:wignerfunctimedeppotappr}
\varrho(t,\vec{x},\vec{p}) \approx {\sf U}(t,t_P,\vec{p})  g(t,\vec{x},\vec{p}) {\sf U}^\dagger(t,t_P,\vec{p}) \,,
\end{align}
with the shape factor given by
\begin{align}
\label{eq:shapefactconstpotlowestord}
g(t,\vec{x},\vec{p}) & \approx \int\frac{d^3\vec{\Delta}}{(2\pi)^3} e^{-i\vec{\Delta}(\vec{v}_\vec{p}(t-t_P)-\vec{x})}
\psi(t_P,\vec{p}+\textstyle{\frac12}\vec{\Delta}) \psi^\dagger(t_P,\vec{p}-\textstyle{\frac12}\vec{\Delta})\,.
\end{align}
Whereas the phase factor ${\sf U}(t,t_P,\vec{p})$ (the terms \emph{phase factor} and \emph{evolution operator} are used here interchangeably) depends on the matter profile along the neutrino trajectory, the shape factor is determined (in the ultrarelativistic limit, see reference \citep{Kartavtsev:2014mea} for an analysis of subleading corrections) only by the initial conditions.

For the initial conditions specified in equation \eqref{eq:gaussianwp} the flavor-basis counterpart of the shape factor, $g_{\alpha \beta} = U_{\alpha i}\, g_{ij}\, U^\dagger_{j\beta}$, reads \citep{Kartavtsev:2014mea}  
\begin{align}
\label{eq:shapewignergaussian}
g_{\alpha\beta}(t,\vec{x},\vec{p}) = \delta_{\alpha e}\delta_{e\beta}\cdot 2^3\exp\biggl(-\frac{(\vec{p}-\vec{p}_w)^2}{2\sigma_p^2}\biggr)\,
\exp\biggl(-\frac{(\vec{v}_\vec{p}(t-t_P)-(\vec{x}-\vec{x}_P))^2}{2\sigma^2_x}\biggr)\,,
\end{align}
where $\sigma_x$ is the coordinate uncertainty defined by $\sigma_x\sigma_p=\frac12$. At $t=t_P$ the shape factor is centered in the vicinity of the production point $\vec{x}_P$. Note that $\int d^3\vec{x} \int d^3\vec{p}/(2\pi)^3 g_{\alpha\beta}(t,\vec{x},\vec{p}) = \delta_{\alpha e}\delta_{e\beta}$, i.e. the shape factor is normalized to unity.

An experiment detecting neutrinos via the charged-current interactions \citep{Abe:2016nxk,Alimonti:2008gc} is sensitive to the individual components of the flavor-basis Wigner function, $\varrho_{\alpha\alpha} = U_{\alpha i} \, \varrho_{ij}\, U^\dagger_{j\alpha}$. For the initial conditions specified in equation \eqref{eq:shapewignergaussian} we find	 
\begin{align}
\label{eq:wigfuncee}
\varrho_{\alpha\alpha}(t,\vec{x},\vec{p}) = |{\sf U}_{\alpha e}(t,t_P,\vec{p})|^2\,g_{ee}(t,\vec{x},\vec{p})\,,
\end{align}
where ${\sf U}_{\alpha e} = U_{\alpha i}\, {\sf U}_{ij}\, U^\dagger_{je}$ is the flavor-basis counterpart of the phase factor. On the other hand, an experiment detecting neutrinos via the neutral-current interactions \citep{Ahmad:2002jz} does not distinguish between the neutrino flavors and is sensitive to $\sum_\alpha \varrho_{\alpha\alpha}$. Unitarity of the PMNS matrix and of the phase factor implies that
$\sum_{\alpha}\varrho_{\alpha\alpha}(t,\vec{x},\vec{p}) = g_{ee}(t,\vec{x},\vec{p})$. The ratio 
\begin{align}
P_\alpha(t,\vec{p})  \equiv \varrho_{\alpha\alpha}(t,\vec{x},\vec{p}) / {\textstyle\sum}_{\alpha}\varrho_{\alpha\alpha}(t,\vec{x},\vec{p}) =  |{\sf U}_{\alpha e}(t,t_P,\vec{p})|^2
\end{align}
gives the  probability of detecting at a time $t$ the momentum mode $\vec{p}$ of the neutrino wave packet  in a flavor $\alpha$. Unitarity of the PMNS matrix and of the phase factor implies that $\sum_\alpha P_\alpha(t,\vec{p}) =1$. 

Because energies of solar neutrinos are below the muon and tau production thresholds, in the remainder of this section we will focus on calculating $P_e(t,\vec{p})$ and $\varrho_{ee}(t,\vec{x},\vec{p})$. To emphasize that neutrino propagation in a spatially inhomogeneous potential is approximated here by propagation in a spatially homogeneous but time-dependent potential we will keep the time argument of $P_e$ instead of expressing it through the distance traveled by the neutrino wave packet.

\paragraph{Propagation in the Sun.} The neutral-current neutrino interactions with matter are flavor-diagonal and only induce an overall phase in $\mathsf{U}$, which does not affect $ |{\sf U}_{ee}(t,t_P,\vec{p})|$. The charged-current neutrino interactions with electrons produce a potential $\mathsf{V}_e = \sqrt{2}G_F n_e$, where $n_e$ is the electron number density \citep{Giunti:2007ry}. The latter is predicted by the Standard Solar Model and is well approximated by $n_e/N_A = 245\,\exp(-10.54\,R/R_\odot)$ \citep{Bahcall:2000nu}, with $N_A$ being the Avogadro constant. The resulting Hamiltonian is given in the mass basis by ${\sf H}_{ij}(t,\vec{p}) = \delta_{ij}\,\omega_i(\vec{p}) + U^\dagger_{ie} {\sf V}_e(t) U_{ej}$. In the standard parametrization the PMNS matrix reads \citep{PhysRevD.98.030001}
\begin{align}
U(\theta_{23},\theta_{13},\theta_{12},\delta)&=O_{23}(\theta_{23})\Gamma_\delta O_{13}(\theta_{13}) \Gamma^\dagger_\delta O_{12}(\theta_{12})\,,
\end{align}
where $O_{ij}$ is the orthogonal rotation matrix in the $ij$-plane which depends on the mixing angle $\theta_{ij}$, and $\Gamma_\delta = {\rm diag}(1, 1, e^{i\delta})$, where  $\delta$ is the Dirac-type CP-violating phase. Following the simple yet accurate approximation scheme developed in reference \citep{Akhmedov:2004rq} we neglect the $(1,3)$, $(3,1)$, $(2,3)$ and $(3,2)$ elements of the Hamiltonian, see appendix \ref{sec:apprthreeflavor} for more details. This yields
\begin{align}
{\sf H}(t,\vec{p}) \approx 
\begin{pmatrix}
\omega_1(\vec{p})+{\sf V}_e(t)\,c^2_{13} c^2_{12} & {\sf V}_e(t)\,c^2_{13} c_{12}s_{12} & 0 \\ 
{\sf V}_e(t)\,c^2_{13} c_{12}s_{12}  & \omega_2(\vec{p}) + {\sf V}_e(t)\,c^2_{13} s^2_{12}  & 0\\
0  & 0 & \omega_3(\vec{p})+{\sf V}_e(t)\,s^2_{13}
\end{pmatrix} \,.
\end{align}
This Hamiltonian is diagonalized by an orthogonal transformation in the 12-plane with the rotation angle $\vartheta_{12}$ determined by 
\begin{align}
\label{eq:diagonalizationangle}
\tan 2\vartheta_{12}(t,\vec{p}) = \frac{{\sf V}_e(t) c^2_{13}\cdot\sin{2\theta_{12}}}{\Delta\omega_{21}(\vec{p})-{\sf V}_e(t) c^2_{13}\cdot\cos{2\theta_{12}}}\,.
\end{align}
For $\Delta\omega_{21}(\vec{p}) = {\sf V}_e(t) c^2_{13}\cdot\cos{2\theta_{12}}$ mixing of the neutrino mass eigenstates is maximal, $\vartheta_{12}\rightarrow \pi/4$, i.e. we encounter the celebrated MSW resonance \citep{Wolfenstein:1977ue,Mikheev:1986gs}. The energies of the propagation eigen\-states are given by
$\mathsf{E}=O^\dagger_{12}(\vartheta_{12})\mathsf{H} O_{12}(\vartheta_{12})  = diag (\bar{E}-\Delta E/2,\bar{E}+\Delta E/2,E_3)$, where
\begin{align}
\Delta E & = \bigl[(\Delta \omega_{21}(\vec{p})-{\sf V}_e(t) c^2_{13} \cdot \cos 2\theta_{12})^2 + ({\sf V}_e(t) c^2_{13} \cdot \sin 2\theta_{12})^2\bigr]^\frac12\,.
\end{align}
Note that $\Delta E$ reaches its minimal yet nonzero value at the MSW resonance. This effect is known as avoided level crossing \citep{PhysRevD.22.2718,Dighe:1999bi}. Some further details can be found in appendix \ref{sec:apprthreeflavor}.

Neutrino propagation in the Sun is very close to adiabatic \citep{Smirnov:2003da,deHolanda:2003nj,deHolanda:2003tx,deHolanda:2004fd,Ioannisian:2004jk,Goswami:2004cn,Blennow:2013rca,Maltoni:2015kca,Smirnov:2016xzf}. In the propagation basis the evolution operator is given in the adiabatic limit by $\exp[-i\phi(t_S,t_P,\vec{p})]$, where 
\begin{align}
\phi(t_S,t_P,\vec{p}) \equiv \int^{t_S}_{t_P}{\sf E}(\tau,\vec{p})d\tau\,
\end{align}
and $t_S$ is the time when the neutrino reaches the solar surface. Projecting the propagation eigenstates on the mass eigenstates at $t_P$ and $t_S$ we arrive at the evolution operator in the mass basis,
\begin{align}
\label{eq:adiabaticevol}
{\sf U}(t_S,t_P,\vec{p}) = O_{12}(\vartheta_{12}(t_S,\vec{p})) \exp[-i\phi(t_S,t_P,\vec{p})] O^\dagger_{12}(\vartheta_{12}(t_P,\vec{p}))\,.
\end{align}
A formal derivation of equation \eqref{eq:adiabaticevol} is presented in appendix \ref{sec:adiabaticprop}.

At the solar surface the matter density is equal to zero and the Hamiltonian is diagonal in the mass basis, therefore $\vartheta_{12}(t_S,\vec{p}) = 0$. For low-energy neutrinos $\Delta \omega_{21}(\vec{p}) \gg {\sf V}_e(t) c^2_{13} \cdot \cos 2\theta_{12}$ and hence $\vartheta_{12}(t_P,\vec{p})$ is close to zero as well. That is, the matter effects are small and the oscillations proceed as in vacuum. On the contrary, for the high-energy neutrinos produced close to the center $\Delta \omega_{21}(\vec{p}) \ll {\sf V}_e(t) c^2_{13} \cdot \cos 2\theta_{12}$ and therefore   $\vartheta_{12}(t_P,\vec{p})\approx \pi/2-\theta_{12}$, see appendix \ref{sec:apprthreeflavor} for more details. For the electron-neutrino initial conditions, see equations \eqref{eq:gaussianwp} and \eqref{eq:shapewignergaussian}, in the limit $\theta_{13}\rightarrow 0$ the resulting Wigner function reads in the mass basis $\varrho_{ij}(t_S,\vec{x},\vec{p})= \delta_{i2}\delta_{2j}\, g_{ee}(t_S,\vec{x},\vec{p})$. Thus, for $\theta_{13}\rightarrow 0$ this fraction of solar neutrinos is adiabatically converted into the second mass eigenstate by the matter effects. For a non-zero $\theta_{13}$ we obtain $\varrho_{22} \propto c^2_{13}$ and $\varrho_{33} \propto s^2_{13}$, while $\varrho_{23}$ and $\varrho_{32}$ are proportional to $c_{13}s_{13}$. Hence, to leading order in $\theta_{13}$ this fraction of solar neutrinos leaves the Sun as a superposition of the second and third mass eigenstates. 

\paragraph{Day-time neutrinos.} Using the composition property we can represent the evolution operator for a neutrino reaching the Earth surface during day-time in the form 
\begin{align}
\label{eq:Umassday}
{\sf U}(t_D,t_P,\vec{p}) = {\sf U}(t_D,t_S,\vec{p})  {\sf U}(t_S,t_P,\vec{p})\,,
\end{align}
where $t_D$ is the moment when the neutrino reaches the Earth. Because in vacuum the Hamiltonian is diagonal in the mass-basis,  ${\sf U}(t_D,t_S,\vec{p})=\exp[-i\phi(t_D,t_S,\vec{p})]$ is diagonal as well. Substituting equation \eqref{eq:adiabaticevol} into equation \eqref{eq:Umassday} we obtain 
\begin{align}
\label{eq:Udayneutrinos}
{\sf U}(t_D,t_P,\vec{p}) = \exp[-i\phi(t_D,t_P,\vec{p})] O^\dagger_{12}(\vartheta_{12}(t_P,\vec{p}))\,.
\end{align}
Because two subsequent rotations, by $\vartheta_{12}(t_P,\vec{p})$ and by $\theta_{12}$, in the 12-plane combine to a rotation by $\theta^P_{12} \equiv \theta_{12}(t_P,\vec{p}) = \theta_{12}+\vartheta_{12}(t_P,\vec{p})$ we obtain in the flavor basis
\begin{align}
{\sf U}_{ee}(t_D,t_P,\vec{p}) = U_{ei}(\theta_{23},\theta_{13},\theta_{12},\delta)
\exp[-i\phi_i(t_D,t_P,\vec{p})]
U^\dagger_{ie}(\theta_{23},\theta_{13},\theta^P_{12},\delta)\,.
\end{align}
Its absolute square contains non-oscillating contributions,
\begin{align}
\label{eq:Udaynonosc}
P_{e}(t_D,\vec{p}) \ni \sum_i |U_{ei}(\theta_{23},\theta_{13},\theta_{12},\delta)|^2
|U^\dagger_{ie}(\theta_{23},\theta_{13},\theta^P_{12},\delta)|^2 \,,
\end{align}
as well as oscillating terms with phases $e^{-i\Delta\phi_{ij}(t_D,t_P,\vec{p})}$. 

The collision integral that describes neutrino interactions in the detector contains an integration over the neutrino momentum, $\int d^3\vec{k}/(2\pi)^3 f(\vec{p},\vec{k})\varrho_{ee}(t,\vec{x},\vec{k})$, where $f(\vec{p},\vec{k})$ (schematically) models the detector's momentum resolution function. In particular, the momentum resolution function filters out those neutrinos whose momentum lies below the detection threshold. Momentum resolution of realistic detectors is far below $\sigma_p$. Hence, the resolution function is practically constant in the range $\sigma_p$ of momenta around the characteristic momentum $\vec{p}_w$ and the integral can be approximated by $f(\vec{p},\vec{p}_w)\int d^3\vec{k}/(2\pi)^3 \varrho_{ee}(t,\vec{x},\vec{k})$. The remaining integration over $\vec{k}$ yields the neutrino density matrix $\rho_{ee}(t,\vec{x})$. Taking into account that $\vartheta_{12}(t,\vec{p})$ weakly depends on the $\sim \sigma_p$ deviations of $\vec{p}$ from $\vec{p}_w$ we obtain for the contribution of the non-oscillating terms, see equation \eqref{eq:Udaynonosc}, 
\begin{align}
\label{eq:daynonosc}
\rho_{ee}(t_D,\vec{x}) \ni \sum_i |U_{ei}(\theta_{23},\theta_{13},\theta_{12},\delta)|^2
|U^\dagger_{ie}(\theta_{23},\theta_{13},\theta^P_{12},\delta)|^2 \cdot g_{ee}(t_D,\vec{x})\,,
\end{align}
where now $\theta^P_{12} = \theta_{12}+\vartheta_{12}(t_P,\vec{p}_w)$, and 
\begin{align}
g_{ee}(t,\vec{x}) &  = \left(2\pi\sigma^2_x\right)^{-\frac32} \exp\biggl(-\frac{(\vec{v}_{\vec{p}_w}(t-t_P)-(\vec{x}-\vec{x}_P))^2}{2\sigma^2_x}\biggr)\,.
\end{align}
To estimate the contribution of the oscillating terms we expand $\Delta\phi_{ij}(t_D,t_P,\vec{p})$ around the characteristic mo\-mentum $\vec{p}_w$, $\Delta\phi_{ij}(t_D,t_P,\vec{p}) \approx \Delta\phi_{ij}(t_D,t_P,\vec{p}_w) + \partial_\vec{p}\Delta\phi_{ij}(t_D,t_P,\vec{p}_w)(\vec{p}-\vec{p}_w)$. In this approximation and for Gaussian initial conditions the momentum integration can be done analytically and yields terms proportional to 
\begin{align}
\label{eq:dayosc}
\rho_{ee}(t_D,\vec{x}) \ni e^{-i\Delta\phi_{ij}(t_D,t_P,\vec{p}_w)}\,
\exp\left(-\frac{(\partial_\vec{p}\Delta\phi_ {ij}(t_D,t_P,\vec{p}_w))^2}{8\sigma^2_x}\right)\cdot g_{ee}(t,\vec{x})\,.
\end{align}
Because the distance from the Sun to the  Earth is more than 200 times larger than the solar radius, the phase difference $\Delta\phi_ {ij}(t_D,t_P,\vec{p}) = \Delta\phi_ {ij}(t_D,t_S,\vec{p}) + \Delta\phi_ {ij}(t_S,t_P,\vec{p})$ is dominated by the first term, $\Delta\phi_ {ij}(t_D,t_S,\vec{p}) = \Delta\omega_{ij}(\vec{p})(t_D-t_S)$. For low energy neutrinos the matter effects are small and the second term is also well approximated by $\Delta\phi_ {ij}(t_S,t_P,\vec{p}) = \Delta\omega_{ij}(\vec{p})(t_S-t_P)$. On the other hand, the matter effects can  affect  $\Delta\phi_ {12}(t_S,t_P,\vec{p})$ for high-energy neutrinos, see appendix \ref{sec:kindecoherday} for more details. However, the first term alone is sufficient to ensure that $|\partial_\vec{p}\Delta\phi_ {ij}(t_D,t_P,\vec{p}_w)| \gg \sigma_x$. Hence, the right-hand side of equation \eqref{eq:dayosc} is exponentially suppressed. Thus, in a (realistic) detector unable to resolve momenta that differ by less than $\sigma_p$ the oscillating terms average to zero. This is a manifestation of the effect of kinematic decoherence by wave packet separation \citep{Akhmedov:2009rb}.

\begin{figure}[t!]
\center
\includegraphics[width=0.495\textwidth]{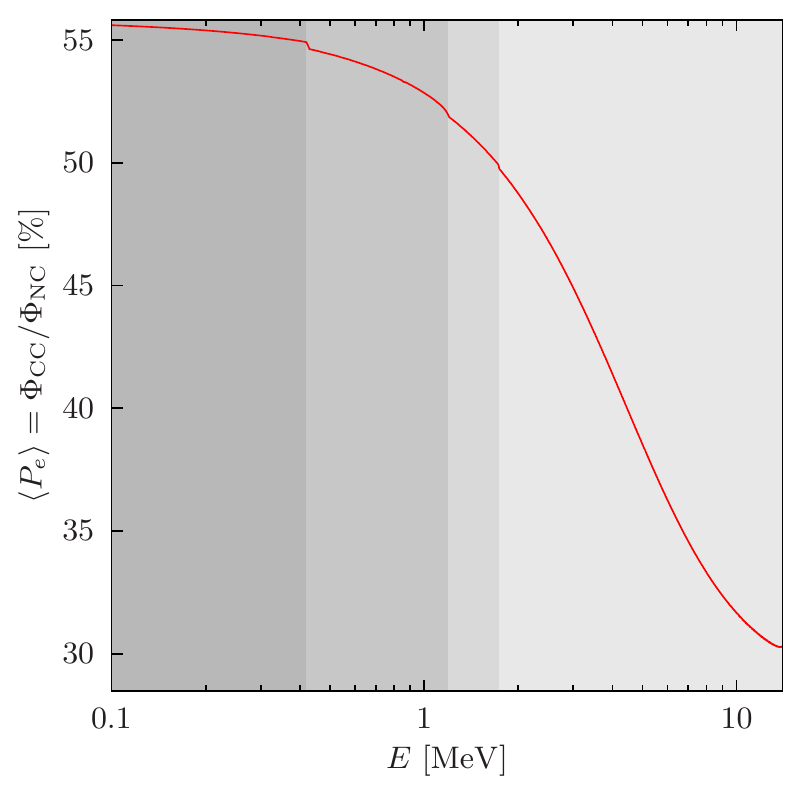}
\includegraphics[width=0.497\textwidth]{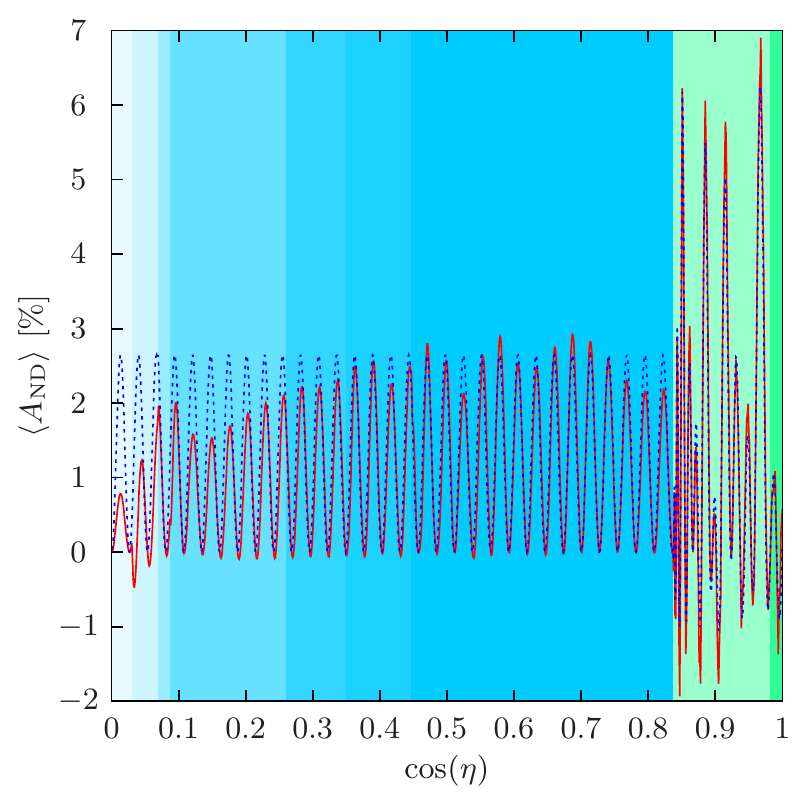}
\caption{\label{fig:daynightneutrinos} \emph{Left panel:} Energy dependence of the survival probability for the day-time neutrinos (averaged over the neutrino production chains and regions).  The gray shaded areas mark energy intervals dominated by a particular flux component and match those in figure \ref{fig:spectrum}. \emph{Right panel:} The red curve depicts the night-day asymmetry (averaged over the neutrino production chains and regions) as function of the nadir angle $\eta$ for the neutrino energy $E=10$ MeV calculated using the PREM model \citep{DZIEWONSKI1981297}. The dashed blue curve denotes the night-day asymmetry computed in the two layer (mantle-core) approxi\-mation. The blue-shaded areas correspond to the trajectories crossing only the mantle, whereas the green-shaded areas correspond to the trajectories crossing also the core. Each shaded region marks one of the nine layers of the PREM model.}
\end{figure}

These arguments imply that it is sufficient to consider only the contribution in equation \eqref{eq:daynonosc}. The latter can be written in the form $\rho_{ee}(t_D,\vec{x}) = P_{e}(t_D)\, g_{ee}(t_D,\vec{x})$,  where $P_{e}(t_D)$ is the survival probability of the electron neutrino. Expressed in terms of the (effective) mixing angles it reads
\begin{align}
\label{eq:Pday}
P_{e}(t_D)  = c^4_{13}\bigl[c^2_{12}\cos^2(\theta^P_{12})+s^2_{12}\sin^2(\theta^P_{12})\bigr]+s^4_{13}\,.
\end{align} 
Note that the  right-hand side of equation \eqref{eq:Pday} depends on the conditions at the production point via $\theta^P_{12}$ but does not depend on $t_D$. Hence, as a result of wave packet separation, flavor composition of the day-time neutrinos does not exhibit a seasonal variation. For the low energy solar neutrinos the matter effects are small and $\theta^P_{12} = \theta_{12}$, hence $P_e = c^4_{13}\bigl[c^4_{12}+s^4_{12}\bigr]+s^4_{13} \approx 0.56$. On the other hand, for the high-energy fraction $\theta^P_{12} = \pi/2$ and therefore  $P_e= c^4_{13}s^2_{12}+s^4_{13} \approx 0.28$. This observation explains the maximal and minimal values of $P_e$ in the left panel of figure \ref{fig:daynightneutrinos}.

\paragraph{Night-time neutrinos.} Crossing the Earth surface, the mass eigenstates split into the matter propagation eigenstates and start to oscillate \citep{Maltoni:2015kca}. Using the composition property we can represent the respective evolution operator as 
\begin{align}
{\sf U}(t_N,t_P,\vec{p})={\sf U}(t_N,t_D,\vec{p}) {\sf U}(t_D,t_P,\vec{p})\,,
\end{align}
where $t_N$ is the moment when the neutrino reaches the detector after crossing the Earth. Recalling equation \eqref{eq:Udayneutrinos}, we obtain in the flavor basis
\begin{align}
{\sf U}_{ee}(t_N,t_P,\vec{p}) & = [U(\theta_{23},\theta_{13},\theta_{12},\delta) {\sf U}(t_N,t_D,\vec{p})]_{ei}\nonumber \\ &\times  \exp[-i\phi_i(t_D,t_P,\vec{p})]U^\dagger_{ie}(\theta_{23},\theta_{13},\theta^P_{12},\delta)\,.
\end{align}
Its absolute square contains terms in which the propagation phases $\phi(t_D,t_P,\vec{p})$ cancel out, 
\begin{align}
P_{e}(t_N,\vec{p}) \ni \sum_i |[U(\theta_{23},\theta_{13},\theta_{12},\delta) {\sf U}(t_N,t_D,\vec{p})]_{ei}|^2|U^\dagger_{ie}(\theta_{23},\theta_{13},\theta^P_{12},\delta)|^2\,,
\end{align}
as well as terms proportional to $e^{-i\Delta\phi_{ij}(t_D,t_P,\vec{p})}$. Because the neutrino trajectory in the Earth is at least four orders of magnitude shorter than the distance between the Sun and the Earth,  $\phi(t_N,t_D,\vec{p})$ is negligibly small compared to $\phi(t_D,t_P,\vec{p})$. Hence, it can be ignored when analyzing the contribution of the $e^{-i\Delta\phi_{ij}(t_D,t_P,\vec{p})}$ terms. As has been argued in the preceding paragraph, the latter are exponentially suppressed due to the effect of wave packet separation and can be neglected. This results in  $\rho_{ee}(t_N,\vec{x}) \approx P_{e}(t_N)\, g_{ee}(t_N,\vec{x})$,  with the survival probability $P_{e}(t_N)$ given by
\begin{align}
\label{eq:Pnight}
P_{e}(t_N) =  \sum_i |[U(\theta_{23},\theta_{13},\theta_{12},\delta) {\sf U}(t_N,t_D,\vec{p}_w)]_{ei}|^2|U^\dagger_{ie}(\theta_{23},\theta_{13},\theta^P_{12},\delta)|^2\,.
\end{align}
In the limit ${\sf U}(t_N,t_D,\vec{p}_w)\rightarrow \mathbb{1}$, where $\mathbb{1}$ is the identity matrix, $P_{e}(t_N)\rightarrow P_{e}(t_D)$. As the Earth matter effects are small, instead of computing $P_{e}(t_N)$ it is convenient to analyze $P_{e}(t_N)-P_{e}(t_D)$ instead. The evolution operator ${\sf U}(t_N,t_D,\vec{p}_w)$ is determined by the matter density profile along the neutrino trajectory. The Preliminary Reference Earth Model (PREM) \citep{DZIEWONSKI1981297} distinguishes nine layers with the matter density slowly varying inside the layers  and sharp density changes at the borders between the layers. The detailed density profile can be found in table II of reference \citep{DZIEWONSKI1981297}. Neutrino propagation within the layers is very close to adiabatic, whereas at the borders the adiabaticity is maximally violated \citep{deHolanda:2004fd,Akhmedov:2004rq,Ioannisian:2017dkx}. Hence, ${\sf U}(t_N,t_D,\vec{p}_w)$ can be computed by stacking up the time evolution operators for the propagation in the individual layers calculated using equation \eqref{eq:evolopadiabatic}. The trajectory, and therefore also the number of the layers crossed, depends on the nadir angle $\eta$, see e.g. figure 1 in reference \citep{Lisi:1997yc}. The night-day asymmetry \citep{Ioannisian:2017dkx,Bakhti:2020tcj}, $A_{ND} = [P_{e}(t_N)-P_{e}(t_D)]/P_{e}(t_D)$, computed numerically as a  function of $\cos(\eta)$ for the neutrino energy $E=10$~MeV is presented in the right panel of figure \ref{fig:daynightneutrinos}. Its shape can be qualitatively understood in the two-layer approximation, that distinguishes only between the mantle and the core. For a neutrino crossing only the mantle ($0 \leq \cos(\eta) \lesssim 0.84$)
\begin{align}
\mathsf{U}(t_N,t_D,\vec{p}) = O_{12}(\vartheta_{12}(t_{N_-},\vec{p})) \exp[-i\phi(t_N,t_D,\vec{p})] O^\dagger_{12}(\vartheta_{12}(t_{D_+},\vec{p}))\,,
\end{align}
where $t_{D_+} = t_{D} + \epsilon$ is the moment shortly after the neutrino enters the Earth, and  $t_{N_-} = t_N-\epsilon$ is the moment shortly before it leaves it. To a good approximation, the matter density profile of the Earth is spherically symmetric, hence $\vartheta_{12}(t_{N_-},\vec{p}) = \vartheta_{12}(t_{D_+},\vec{p})$. The resulting expression for  the difference of the survival probabilities reads \citep{Antonelli:2012qu}
\begin{align}
\label{eq:nightdaymantle}
P_{e}(t_N)-P_{e}(t_D) & =  - c^4_{13}\cos(2\theta^P_{12}) \nonumber\\
& \times \sin(2\vartheta_{12}(t_{D_+},\vec{p}_w))\sin^2(\Delta\phi_{21}(t_N,t_D,\vec{p}_w)/2)\sin(2\theta_{12}(t_{D_+},\vec{p}_w))\,,
\end{align}
see appendix \ref{sec:nightneutrinos} for a few further details. As can be read off from the right panel of figure \ref{fig:daynightneutrinos}, equation \eqref{eq:nightdaymantle} reasonably well reproduces the numerical results for $0.4\lesssim \cos(\eta) \lesssim 0.84$. For a neutrino crossing also the core ($0.84 \lesssim \cos(\eta) \leq 1$)
\begin{align}
\mathsf{U}(t_N,t_D,\vec{p}) & = O_{12}(\vartheta_{12}(t_{N_-},\vec{p})) \exp[-i\phi(t_N,t_2,\vec{p})] O^\dagger_{12}(\vartheta_{12}(t_{2_+},\vec{p}))\nonumber\\
& \times O_{12}(\vartheta_{12}(t_{2_-},\vec{p})) \exp[-i\phi(t_2,t_1,\vec{p})] O^\dagger_{12}(\vartheta_{12}(t_{1_+},\vec{p}))\nonumber\\
& \times O_{12}(\vartheta_{12}(t_{1_-},\vec{p})) \exp[-i\phi(t_1,t_D,\vec{p})] O^\dagger_{12}(\vartheta_{12}(t_{D_+},\vec{p}))\,,
\end{align}
where $t_1$ and $t_2$ (with $t_D < t_1 < t_2 < t_N$) are the moments when the neutrino enters and leaves the core. The approximate spherical symmetry of the Earth implies $\vartheta_{12}(t_{1_-},\vec{p}) = \vartheta_{12}(t_{2_+},\vec{p})$ and $\vartheta_{12}(t_{1_+},\vec{p}) = \vartheta_{12}(t_{2_-},\vec{p})$. The resulting expression for the difference of the survival probabilities reads 
\begin{align}
\label{eq:nightdaycore}
P_{e}(t_N)-P_{e}(t_D) & =  - c^4_{13}\cos(2\theta^P_{12}) \nonumber\\
& \times \bigl(\sin(2\vartheta_{12}(t_{D_+},\vec{p}_w))[\,\sin(\phi(t_1,t_D,\vec{p}_w))\cos(\phi(t_2,t_1,\vec{p}_w)/2) \nonumber\\
&+ \cos(2\Delta\vartheta_{12}(t_1,\vec{p}_w)) \cos(\phi(t_1,t_D,\vec{p}_w))\sin(\phi(t_2,t_1,\vec{p}_w)/2)]\nonumber\\
&+ \cos(2\vartheta_{12}(t_{D_+},\vec{p}_w)) \sin(2\Delta\vartheta_{12}(t_1,\vec{p}_w))\sin(\phi(t_2,t_1,\vec{p}_w)/2) \bigr) \nonumber\\
&\times  \bigl(\sin(2\theta_{12}(t_{D_+},\vec{p}_w))[\,\sin(\phi(t_1,t_D,\vec{p}_w))\cos(\phi(t_2,t_1,\vec{p}_w)/2) \nonumber\\
&+ \cos(2\Delta\vartheta_{12}(t_1,\vec{p}_w)) \cos(\phi(t_1,t_D,\vec{p}_w))\sin(\phi(t_2,t_1,\vec{p}_w)/2)]\nonumber\\
&+ \cos(2\theta_{12}(t_{D_+},\vec{p}_w)) \sin(2\Delta\vartheta_{12}(t_1,\vec{p}_w))\sin(\phi(t_2,t_1,\vec{p}_w)/2) \bigr)\,,
\end{align}
where $\Delta\vartheta_{12}(t_1,\vec{p}) \equiv \vartheta_{12}(t_{1_+},\vec{p}) - \vartheta_{12}(t_{1_-},\vec{p})$ is the jump of the diagonalization angle induced by the jump of the matter density at the border between the mantle and the core, see  appendix \ref{sec:nightneutrinos}. As can be read off from the right panel of figure \ref{fig:daynightneutrinos}, equation \eqref{eq:nightdaycore} acceptably well reproduces the numerical results.

\paragraph{Evolution equation for the single-particle Wigner function.}
As follows from the above considerations, the phenomenology of solar neutrinos is determined by an interplay of the phase and shape factors. The phase factor, equation \eqref{eq:evolutionoperator}, describes the neutrino flavor evolution and satisfies a Schr\"odinger-like equation,
\begin{align}
\label{eq:eqforphasefact}
i\partial_t  {\sf U}(t,t_P,\vec{p}) = {\sf H}(t,\vec{p})  {\sf U}(t,t_P,\vec{p})\,.
\end{align}
The shape factor, equation \eqref{eq:shapefactconstpotlowestord}, describes position of the neutrino wave packet and satisfies the Liouville equation,
\begin{align}
\label{eq:eqforshapefact}
(\partial_t + \vec{v}_\vec{p}\partial_\vec{x})g(t,\vec{x},\vec{p})  = 0 \,.
\end{align}
Combining equations \eqref{eq:wignerfunctimedeppotappr}, \eqref{eq:eqforphasefact} and \eqref{eq:eqforshapefact} we obtain an evolution equation for the Wigner func\-tion (valid in the mass and flavor bases),
\begin{align}
\label{eq:evoleqhompot}
(\partial_t+\vec{v}_\vec{p}\partial_\vec{x})\varrho(t,\vec{x},\vec{p}) = -i[{\sf H}(t,\vec{p}),\varrho(t,\vec{x},\vec{p})]\,.
\end{align}
As has been shown above, supplemented by the initial conditions given (in the flavor basis) by
\begin{align}
\varrho_{\alpha\beta}(t_P,\vec{x},\vec{p})=\delta_{\alpha e}\delta_{\beta e}\, g_{ee}(t_P,\vec{x},\vec{p})
\end{align}
equation \eqref{eq:evoleqhompot} reproduces the standard results for the day-time and night-time neutrinos, including the impact of the effect of wave packet separation on the neutrino flavor composition.  

This implies that solutions of the evolution equation for the Wigner function, supplemented by initial conditions constructed from the neutrino wave function, are consistent with the Heisenberg uncertainty principle. The uncertainty principle is encoded in the shape factor, which is localized neither in coordinate nor in momentum space, see equation \eqref{eq:shapewignergaussian} for an explicit example. Note that unlike the Schr\"odinger equation, equation \eqref{eq:evoleqhompot} does not enforce the uncertainty prin\-ci\-ple and admits also classical solutions describing a particle with a definite coordinate and momentum if supplemented by the respective initial conditions, see Appendix \ref{sec:classicallimit}.

\paragraph{Derivation from the Schr\"odinger equation.} 
The evolution equation \eqref{eq:evoleqhompot} can be derived from the Schr\"odinger equation without resorting to its explicit solutions. To this end we differentiate equation \eqref{eq:wignerfunctionmom} with respect to time and subsequently use equation \eqref{eq:schrodmomentum}. This yields
\begin{align}
\label{eq:evoleqsinglepart1}
\partial_t\varrho_{ij}(t,\vec{x},\vec{p}) & =-i\int\frac{d^3\vec{\Delta}}{(2\pi)^3}e^{i\vec{\Delta}\vec{x}} \nonumber\\
& \times \bigl[ \mathsf{H}_{ik}(t,\vec{p}+\vec{\Delta}/2)\psi_k(t,\vec{p}+\vec{\Delta}/2)\psi^*_j(t,\vec{p}-\vec{\Delta}/2)\nonumber\\
&-\psi_i(t,\vec{p}+\vec{\Delta}/2)\psi^*_k(t,\vec{p}-\vec{\Delta}/2)\mathsf{H}_{kj}(t,\vec{p}-\vec{\Delta}/2)\bigr]\,,
\end{align}
where we have used hermiticity of the Hamiltonian, $\mathsf{H}^*_{jk}(t,\vec{p}-\vec{\Delta}/2)=\mathsf{H}_{kj}(t,\vec{p}-\vec{\Delta}/2)$, in the second line. Expanding the Hamiltonian around $\vec{p}$, \smash{$\mathsf{H}(t,\vec{p}\pm\vec{\Delta}/2)=\sum_n\frac{(\pm\vec{\Delta})^n}{2^n\,n!}\partial^n_\vec{p}\mathsf{H}(t,\vec{p})$}, and using $\vec{\Delta}^n  e^{i\vec{\Delta}\vec{x}}= (-i\partial_\vec{x})^ne^{i\vec{\Delta}\vec{x}}$ we can recast equation \eqref{eq:evoleqsinglepart1} in the form 
\begin{align}
\partial_t\varrho_{ij}(t,\vec{x},\vec{p}) & =-i\sum_n \frac{(-i)^{n}}{2^n\, n!} \bigl[
\partial^n_\vec{p}\mathsf{H}_{ik}(t,\vec{p})\partial^n_\vec{x}\varrho_{kj}(t,\vec{x},\vec{p})\nonumber\\
&-(-1)^n\partial^n_\vec{x}\varrho_{ik}(t,\vec{x},\vec{p})\partial^n_\vec{p}\mathsf{H}_{kj}(t,\vec{p})
\bigr]\,.
\end{align}
In the approximation used above $\partial_\vec{p} {\sf H}\approx \vec{v}_\vec{p}$ and therefore commutes with $\varrho$. Neglecting the $n\geq 2$ terms, i.e. performing the first-order gradient expansion, we recover equation \eqref{eq:evoleqhompot}.

\paragraph{Evolution equation for propagation in an inhomogeneous potential.}
The analysis presented above is based on an approximation replacing neutrino propagation in an inhomogeneous potential by propagation in a homogeneous but time dependent potential. An evolution equation for the Wigner function of a neutrino propagating in a general potential can  be derived directly from the Schr\"odinger equation as well. Conceptually, the derivation is similar to that presented in reference \citep{Kartavtsev:2014mea}, but here we work directly with the Wigner function and use the relativistic approximation, to facilitate comparison with the kinetic equation for the matrix of densities addressed in section \ref{sec:kinapproach}. For $\mathsf{H}_{ij}(t,\vec{x},\vec{p}) = \delta_{ij}\omega_i(\vec{p})+\mathsf{V}_{ij}(t,\vec{x})$ the  momentum-rep\-re\-sen\-ta\-tion coun\-terpart of the Schr\"odinger equation \eqref{eq:schrodeq} reads
\begin{align}
\label{eq:schrodmomentuminhom}
i\partial_t\psi_i(t,\vec{p}) = \int\frac{d^3\vec{k}}{(2\pi)^3}  {\sf H}_{ij}(t,\vec{p},\vec{k})\psi_j(t,\vec{k})\,,
\end{align}
with \cite{Stirner:2018ojk}
\begin{align}
\label{eq:twomomhamiltonian}
{\sf H}_{ij}(t,\vec{p},\vec{k}) = \int d^3\vec{x}\, e^{-i(\vec{p}-\vec{k})\vec{x}}\, {\sf H}_{ij}\bigr(t,\vec{x},{\textstyle\frac{\vec{p}+\vec{k}}{2}}\bigr)\,.
\end{align}
Note that for a spatially homogeneous potential ${\sf H}_{ij}(t,\vec{p},\vec{k}) = (2\pi)^3\delta(\vec{p}-\vec{k}) {\sf H}_{ij}(t,\vec{p})$ and equation \eqref{eq:schrodmomentuminhom} reverts to equation \eqref{eq:schrodmomentum}. The hermiticity condition translates into ${\sf H}_{ij}(t,\vec{p},\vec{k})={\sf H}^*_{ji}(t,\vec{k},\vec{p})$ in the momentum representation.

Differentiating equation \eqref{eq:wignerfunctionmom} with respect to time and using equation \eqref{eq:schrodmomentuminhom} we arrive at 
\begin{align}
\label{eq:timederrhoingom0}
\partial_t\varrho_{ij}(t,\vec{x},\vec{p}) & = -i \int \frac{d^3\vec{\Delta}}{(2\pi)^3} e^{i\vec{\Delta}\vec{x}} \int \frac{d^3\vec{k}}{(2\pi)^3} \nonumber\\
& \times \bigl[\mathsf{H}_{ik}(t,\vec{p}+\vec{\Delta}/2,\vec{k})\psi_k(t,\vec{k})\psi^*_j(t,\vec{p}-\vec{\Delta}/2)\nonumber\\
& - \psi_i(t,\vec{p}+\vec{\Delta}/2)\psi^*_k(t,\vec{k})\mathsf{H}_{kj}(t,\vec{k},\vec{p}-\vec{\Delta}/2)\bigr]\,,
\end{align}
where we have used hermiticity of the Hamiltonian in the third line. Using equation \eqref{eq:twomomhamiltonian} we find
\begin{align}
\label{eq:timederrhoingom}
\partial_t\varrho_{ij}(t,\vec{x},\vec{p}) & = -i \int \frac{d^3\vec{\Delta}}{(2\pi)^3} e^{i\vec{\Delta}\vec{x}} \int \frac{d^3\vec{k}}{(2\pi)^3} \int d^3 \vec{y}\nonumber\\
& \times \bigl[ e^{-i(\vec{p}+\vec{\Delta}/2-\vec{k})\vec{y}}\mathsf{H}_{ik}\bigl(t,\vec{y},{\textstyle\frac{\vec{p}+\vec{\Delta}/2+\vec{k}}{2}}\bigr)\psi_k(t,\vec{k})\psi^*_j(t,\vec{p}-\vec{\Delta}/2)\nonumber\\
& - \psi_i(t,\vec{p}+\vec{\Delta}/2)\psi^*_k(t,\vec{k}) \mathsf{H}_{kj}(t,\vec{y},{\textstyle\frac{\vec{p}-\vec{\Delta}/2+\vec{k}}{2}}\bigr)e^{i(\vec{p}-\vec{\Delta}/2-\vec{k})\vec{y}}\bigr]\,.
\end{align}
Next, we transform the integration variables in the second line of equation \eqref{eq:timederrhoingom} as $\vec{k}\rightarrow\vec{k}+\vec{\Delta}/2$, $\vec{\Delta} \rightarrow \vec{\Delta} + 2(\vec{p}-\vec{k})$, and $\vec{y}\rightarrow\vec{x}+\vec{y}/2$. In its third line we use the transformation $\vec{k}\rightarrow\vec{k}-\vec{\Delta}/2$, $\vec{\Delta} \rightarrow \vec{\Delta} - 2(\vec{p}-\vec{k})$, and $\vec{y}\rightarrow\vec{x}-\vec{y}/2$. The Jacobians of both transformations are equal to unity. These transformations result in 
\begin{align}
\label{eq:timederrhoingom1}
\partial_t\varrho_{ij}(t,\vec{x},\vec{p}) & = -i \int \frac{d^3\vec{\Delta}}{(2\pi)^3} e^{i\vec{\Delta}\vec{x}} \int \frac{d^3\vec{k}}{(2\pi)^3} \int d^3 \vec{y} e^{-i(\vec{p}-\vec{k})\vec{y}}\nonumber\\
& \times \bigl[\mathsf{H}_{ik}\bigl(t,\vec{x}+\vec{y}/2, \vec{p}+\vec{\Delta}/2\bigr)\psi_k(t,\vec{k}+\vec{\Delta}/2)\psi^*_j(t,\vec{k}-\vec{\Delta}/2)\nonumber\\
& - \psi_i(t,\vec{k}+\vec{\Delta}/2)\psi^*_k(t,\vec{k}-\vec{\Delta}/2) \mathsf{H}_{kj}(t,\vec{x}-\vec{y}/2,\vec{p}-\vec{\Delta}/2\bigr)\bigr]\,.
\end{align}
Next, we expand the Hamiltonian, $\mathsf{H}(t,\vec{x}\pm\vec{y}/2,\vec{p}\pm\vec{\Delta}/2)=\sum_{m,n}\frac{(\pm\vec{y})^m}{2^m\,m!}\frac{(\pm\vec{\Delta})^n}{2^n\,n!}\partial^m_\vec{x}\partial^n_\vec{p}\mathsf{H}(t,\vec{x},\vec{p})$. The coefficients $\partial^m_\vec{x}\partial^n_\vec{p}\mathsf{H}(t,\vec{x},\vec{p})$ can be pulled out of the integral. 
Using $\vec{\Delta}^n  e^{i\vec{\Delta}\vec{x}}= (-i\partial_\vec{x})^ne^{i\vec{\Delta}\vec{x}}$ as well as 
$\vec{y}^m  e^{-i\vec{p}\vec{y}}= (i\partial_\vec{p})^me^{-i\vec{p}\vec{y}}$, and integrating over $\vec{y}$ and $\vec{k}$ we obtain
\begin{align}
\label{eq:timederrhoingom2}
\partial_t\varrho_{ij}(t,\vec{x},\vec{p}) & =-i\sum_{n,m} \frac{(-i)^{n-m}}{2^{n+m}\, n!\,m!} \bigl[
\partial^m_\vec{x}\partial^n_\vec{p}\mathsf{H}_{ik}(t,\vec{x},\vec{p})\partial^m_\vec{p}\partial^n_\vec{x}\varrho_{kj}(t,\vec{x},\vec{p}) \nonumber\\
&-(-1)^{n+m}\partial^m_\vec{p}\partial^n_\vec{x}\varrho_{ik}(t,\vec{x},\vec{p})\partial^m_\vec{x}\partial^n_\vec{p}\mathsf{H}_{kj}(t,\vec{x},\vec{p})
\bigr]\,.
\end{align}
Performing the first-order gradient expansion, i.e. keeping at most the first-order derivatives with respect to $\vec{x}$ or $\vec{p}$, we can recast equation \eqref{eq:timederrhoingom2} in the familiar form \citep{Kartavtsev:2014mea},
\begin{align}
\label{eq:timederrhoingom3}
\partial_t\varrho(t,\vec{x},\vec{p}) & + {\textstyle \frac12}\{\partial_\vec{p} \mathsf{H}(t,\vec{x},\vec{p}),\partial_\vec{x}\varrho(t,\vec{x},\vec{p})\} - {\textstyle \frac12}\{\partial_\vec{x} \mathsf{H}(t,\vec{x},\vec{p}),\partial_\vec{p}\varrho(t,\vec{x},\vec{p})\}\nonumber\\
& = -i [ \mathsf{H}(t,\vec{x},\vec{p}),\varrho(t,\vec{x},\vec{p})]\,.
\end{align}
In appendix \ref{sec:densitymatrix} the presented derivation is generalized to states described by the density matrix. In the approximation $\partial_\vec{p} {\sf H}\approx \vec{v}_\vec{p}$ used above the first anticommutator on its left-hand side simplifies to $\vec{v}_\vec{p} \partial_\vec{x}\varrho(t,\vec{x},\vec{p})$ thereby reproducing the velocity term of the Liouville operator in equation \eqref{eq:evoleqhompot}.  A detailed analysis of the Liouville term can be found in reference \citep{Kartavtsev:2014mea}. Let us emphasize, that the presented derivation of the evolution equation does not rely in any way on coarse-graining of the Wigner function \citep{1940264}. The well-known consequence is that the Wigner function is not guaranteed to be non-negative. 

The Wigner function is mathematically well defined for arbitrary wave functions. Because any wave function satisfies the Heisenberg principle, so does also the Wigner function, constructed from the former. On other hand, the applicability of Wigner approach is not guaranteed for arbitrary wave functions. In particular, if the typical space size $\sigma_x$ of the wave function is comparable to its de~Broglie wavelength $\lambda = p^{-1}$, then  it its momentum-space width $\sigma_p$ is comparable to the neutrino momentum. In this case the on-shell approximation, used to derive the evolution equation for the Wigner function, breaks down \cite{Kartavtsev:2014mea} and recasting quantum dynamics in terms of the Wigner function makes little physical sense.  An additional important scale that must be taken into account is the scale of variation of the matter potential. The faster it varies in space and time, the more terms on the right-hand side of equation (2.46) must be kept to provide a sufficiently accurate description of the neutrino propagation. For potentials varying on the scale comparable to or smaller than the neutrino de~Broglie wavelength the use of the Wigner approach also makes little physical sense.

\paragraph{Integration over neutrino spectrum and production points.}
So far we have studied evolution of individual neutrinos. In the detector contributions of neutrinos produced in the dif\-fer\-ent production chains $i$ and different distances $R$ from the center add up. In an experiment detecting neutrinos via the neutral-current interactions the count rate  in an energy bin $[E,E+\Delta E]$ is proportional to the total neutrino flux in this energy range,
\begin{align}
\label{eq:NCcountrate}
\left. \frac{dn}{dt}\right|_{NC}  \propto \sum_i \int^{E+\Delta E}_{E} dE \int^{R_\odot}_0 dR\, \frac{d^2\Phi_i}{dE dR}\,.
\end{align}  
In an experiment detecting neutrinos via the charged-current interactions the differential neutrino flux is weighted by the survival probability $P_e$ and the count rate is proportional to 
\begin{align}
\label{eq:CCcountrate}
\left. \frac{dn}{dt}\right|_{CC}  \propto \sum_i  \int^{E+\Delta E}_{E} dE \int^{R_\odot}_0 dR\, \frac{d^2\Phi_i}{dE dR}\,P_e\,.
\end{align} 
Taking the ratio of equations \eqref{eq:CCcountrate} and \eqref{eq:NCcountrate} for the day-time neutrinos we obtain the average survival probability $\langle P_e \rangle$ depicted in the left panel of figure \ref{fig:daynightneutrinos}. Using equ\-ation \eqref{eq:CCcountrate}  to compute the count rates of the day-time and night-time neutrinos we find the average night-day asymmetry $\langle A_{ND}\rangle$ depicted in the right panel of figure \ref{fig:daynightneutrinos}.

\paragraph{Two sources of kinematic decoherence.}
The count rates can also be expressed directly in terms of the single-par\-ticle Wigner functions. In an experiment detecting neutrinos via the charged-cur\-rent interactions the count rate is proportional to 
\begin{align}
\label{eq:CCcountrate1}
\left. \frac{dn}{dt}\right|_{CC}  \propto \sum_l \int_P d^3\vec{p} \int_V d^3\vec{x} \, \varrho^{(l)}_{ee}(t,\vec{x},\vec{p})\,,
\end{align}  
where $l$ enumerates the solar neutrinos produced by the detection time $t$, $V$ is the detector volume, and $P$ is the momentum-space volume containing the momenta corresponding to the energy range $[E,E+\Delta E]$. The volume integral selects from the sum over $l$ only those neutrinos whose wave packets are located inside the detector at $t$. The wave packets corresponding to different $l$ have different characteristic momenta $\vec{p}_w$, whose distribution reflects the neutrino energy spectrum, see figure \ref{fig:spectrum}. The order of the integrations and summations in equation \eqref{eq:CCcountrate1} is chosen so as to compute first the contribution of each individual neutrino and then sum over the neutrino ensemble. In an experiment sensitive to the neutral-current interactions the count rate is proportional to 
\begin{align}
\label{eq:NCcountrate1}
\left. \frac{dn}{dt}\right|_{NC}  \propto  \sum_l  \sum_\alpha \int_P d^3\vec{p} \int_V d^3\vec{x}\, \varrho^{(l)}_{\alpha\alpha}(t,\vec{x},\vec{p})\,.
\end{align}  

As has been discussed above, if the momentum integration volume $P$, determined by the ex\-per\-i\-men\-tal setup, exceeds the size of the neutrino wave packet in the momentum space, characterized by $\sigma_p$, then the wave packet separation leads to averaging out of the oscillations in the density matrix. As a result, flavor composition of the day-time neutrinos does not exhibit a seasonal variation. This applies even to a hypothetical experiment detecting one neutrino at a time. Crossing the Earth surface the neutrinos start to oscillate. Their path in the Earth is relatively short and in\-suf\-fi\-cient for the wave packet separation to occur. Therefore, the hypothetical experiment detecting one neutrino at a time would observe an oscillation pattern. However, this does not guarantee that the oscillation pattern be observable in a realistic neutrino experiment. At the detection time $t$ flavor of the neutrinos produced at the same time $t_P$ and at the same point $\vec{x}_P$ but possessing different $\vec{p}_w$ differs due to the momentum-dependence of  $\mathsf{U}(t_N,t_D,\vec{p}_w)$ and $\theta^P_{12}$ in equation \eqref{eq:Pnight}. In the sum over $l$ their contributions add-up destructively thereby ``washing out'' the oscillation pattern, unless the detector resolution $\Delta E$ is smaller than the inverse distance from the source to the detector  \cite{Ioannisian:2004jk}. That is, there are two sources of kinematic decoherence, both averaging out the oscillations: dephasing of  different momentum modes of a single-neutrino wave-packet, related to the uncertainty principle, and dephasing of many neutrinos \citep{Stirner:2018ojk}. Technically both are related to the integration over the neutrino momentum and finite momentum resolution of the detector.

Furthermore, even the neutrinos characterized by the same $\vec{p}_w$ but produced in different parts of the Sun arrive to the detector in different flavor states because the effective mixing angle $\theta^P_{12}$ depends on the matter density at the production point, see equation \eqref{eq:diagonalizationangle}.  This effect also contributes to the ``washout" of the oscillation pattern \citep{deHolanda:2004fd}.

Because this section is devoted to the single-particle quantum-mechanical approach to neutrino oscillations, here we intentionally speak of wave packets of individual neutrinos. While this terminology serves the purpose and is sufficient to describe physics of solar neutrinos, it is nevertheless misleading. Neutrinos are quantum-mechanically not distinguishable, and the $N$-neutrino systems discussed in this paragraph are described by a common wave packet. This point is elaborated on in the next section.

\section{\label{sec:manypartqm}$N$-particle quantum-mechanical approach}
Given the low density of solar neutrinos and the resulting smallness of the neutrino-neutrino interactions, the individual neutrinos can be considered as independent. In this approximation wave function of the $N$-particle neutrino system is given by the Slater determinant. The anti-symmetry of the wave function under exchanges of any two particles allows the definition of a reduced single-particle Wigner function, with the coordinates and momenta of $N-1$ particles integrated over. In the limit of non-overlapping individual wave functions the latter is simply the sum of the individual single-particle Wigner functions. Hence, it automatically reproduces the summation prescription used in the single-particle quantum-mechanical approach. An evolution equation for the reduced single-particle Wigner function can be derived from the Schr\"odinger equation. Its form matches the form of the evolution equation for the single-particle Wigner function. Solutions of the evolution equation constructed from solutions of the Schr\"odinger equation are consistent with the Heisenberg and Pauli principles at any given time, in particular at $t=t_P$. Turning this the over way around, a solution of the evolution equation with the initial conditions constructed from a solution of the Schr\"odinger equation is consistent with the Heisenberg and Pauli principles also for  $t > t_P$. On the other hand, unlike the Schr\"odinger equation, the evolution equation also admits classical solutions that describe an ensemble of particles with definite coordinates and momenta, if supplemented by the classical initial conditions.

\paragraph{$N$-particle wave function.}
Due to the low density of solar neutrinos, their self-interactions can be neglected. Therefore, the individual neutrinos can be considered independent. In this ap\-prox\-i\-mation wave function of the $N$-particle neutrino system, that by the Pauli exclusion principle must be antisymmetric under a permutation of any two particles, is given by the Slater  de\-ter\-minant. In the mass basis
\begin{align}
\label{eq:Npartwavefunc}
\psi_{i_1\ldots i_N}(t,\vec{p}_1,\ldots, \vec{p}_N)= {\mathcal N}
\begin{vmatrix}
\psi_{1,\,i_1}(t,\vec{p}_1) & \ldots & \psi_{1,\,i_N}(t,\vec{p}_N) \\
\ldots & \ldots & \ldots \\
\psi_{N,\,i_1}(t,\vec{p}_1) & \ldots & \psi_{N,\,i_N}(t,\vec{p}_N) 
\end{vmatrix}\,.
\end{align}
For non-overlapping individual wave functions the normalization factor is given by ${\mathcal N}=(N!)^{-\frac12}$. For overlapping wave functions, $\sum_i\int d^3\vec{p}/(2\pi)^3 \psi_{a,i}(t,\vec{p})\psi^*_{b,i}(t,\vec{p}) \neq 0$ for $a\neq b$, the overlap contribution  must be included into the normalization. One can show that for identical particles the overlap contribution is time-independent and so is the normalization factor, as expected. Note that a set of orthogonal wave functions can be constructed from $\psi_{a,i}$ using the Gram-Schmidt method. The flavor-basis counterpart of equation \eqref{eq:Npartwavefunc} is obtained by replacing the mass-basis wave functions by the flavor-basis ones,  $\psi_{\alpha_1\ldots \alpha_N}(t,\vec{p}_1,\ldots, \vec{p}_N) = U_{\alpha_1 i_1}\ldots U_{\alpha_N i_N}\, \psi_{i_1\ldots i_N}(t,\vec{p}_1,\ldots, \vec{p}_N)$. 
 
Differentiating equation \eqref{eq:Npartwavefunc} with respect to time and applying equation \eqref{eq:schrodmomentuminhom}, we obtain the $N$-particle generalization of the momentum-representation Schr\"odinger equation,
\begin{align}
\label{eq:NparticleSchrodinger}
i\partial_t \psi_{i_1\ldots i_N}(t,\vec{p}_1,\ldots, \vec{p}_N) & = \int\frac{d^3\vec{k}_1}{(2\pi)^3}\ldots \frac{d^3\vec{k}_N}{(2\pi)^3}\, {\sf H}_{i_1 \ldots i_N\, j_1\ldots j_N}(t,\vec{p}_1,\ldots,\vec{p}_N,\vec{k}_1,\ldots,\vec{k}_N)\nonumber\\
&\times \psi_{j_1\ldots j_N}(t,\vec{k}_1,\ldots, \vec{k}_N)\,.
\end{align}
As expected for a system of independent particles, the $N$-particle Hamiltonian is given by the sum of the individual Hamiltonians, 
\begin{align}
\label{eq:NparticleHamiltonian}
{\sf H}_{i_1 \ldots i_N\, j_1\ldots j_N}(t,\vec{p}_1,\ldots,\vec{p}_N,\vec{k}_1,\ldots,\vec{k}_N) & = {\sf H}_{i_1j_1}(t,\vec{p}_1,\vec{k}_1) \ldots  \delta_{i_Nj_N}(2\pi)^3\delta(\vec{p}_N-\vec{k}_N) \nonumber\\
& +  \ldots \nonumber \\
& +  \delta_{i_1j_1} (2\pi)^3\delta(\vec{p}_1-\vec{k}_1)\ldots{\sf H}_{i_Nj_N}(t,\vec{p}_N,\vec{k}_N)\,.
\end{align}
From hermiticity of the single-particle Hamiltonians, ${\sf H}_{i_1j_1}(t,\vec{p}_1,\vec{k}_1) = {\sf H}^*_{j_1i_1}(t,\vec{k}_1,\vec{p}_1)$, it follows that ${\sf H}_{i_1 \ldots i_N\, j_1\ldots j_N}(t,\vec{p}_1,\ldots,\vec{p}_N,\vec{k}_1,\ldots,\vec{k}_N)={\sf H}^*_{j_1 \ldots j_N\, i_1\ldots i_N}(t,\vec{k}_1,\ldots,\vec{k}_N,\vec{p}_1,\ldots,\vec{p}_N)$.

\paragraph{Reduced single-particle Wigner function.}
The Wigner function of the $N$-particle system is de\-fined analogously to the single-particle one \citep{Hillery:1983ms},
\begin{align}
\label{eq:NparticleWigner}
\varrho_{i_1\ldots i_N j_1\ldots j_N}(t,\vec{x}_1,\ldots,\vec{x}_N,\vec{p}_1,\ldots,\vec{p}_N)& = \int \frac{d^3\vec{\Delta}_1}{(2\pi)^3}\ldots \frac{d^3\vec{\Delta}_N}{(2\pi)^3}\, e^{i\sum_l \vec{\Delta}_l \vec{x}_l} \nonumber\\
& \times \psi_{i_1 \ldots i_N}(t,\vec{p}_1+\vec{\Delta}_1/2,\ldots,\vec{p}_N+\vec{\Delta}_N/2)\nonumber\\
& \times \psi^*_{j_1 \ldots j_N}(t,\vec{p}_1-\vec{\Delta}_1/2,\ldots,\vec{p}_N-\vec{\Delta}_N/2)\,.
\end{align}
The antisymmetry of the wave function under a permutation of any two particles allows the definition of a reduced single-particle Wigner function with $N-1$ coordinates and momenta integrated over \citep{Hillery:1983ms},
\begin{align}
\varrho_{ij}(t,\vec{x},\vec{p}) & = N \sum_{l_2 \ldots l_N}\int d^3\vec{x}_2\ldots d^3\vec{x}_N\int \frac{d^3\vec{p}_2}{(2\pi)^3}\ldots \frac{d^3\vec{p}_N}{(2\pi)^3}\nonumber\\
&\times  \varrho_{il_2\ldots l_N\,jl_2\ldots l_N}(t,\vec{x},\vec{x}_2,\ldots,\vec{x}_N,\vec{p},\vec{p}_2,\ldots,\vec{p}_N)\,.
\end{align}
Expressed in terms of the momentum-representation, $N$-particle wave function it reads,
\begin{align}
\label{eq:redwigfunc}
\varrho_{ij}(t,\vec{x},\vec{p}) & = N \int \frac{d^3\vec{\Delta}}{(2\pi)^3} e^{i\vec{\Delta}\vec{x}} \sum_{l_2\ldots l_N} \int \frac{d^3\vec{p}_2}{(2\pi)^3} \ldots \frac{d^3\vec{p}_N}{(2\pi)^3}\nonumber\\
& \times \psi_{il_2\ldots l_N}(t,\vec{p}+\vec{\Delta}/2,\vec{p}_2,\ldots,\vec{p}_N) \psi^*_{jl_2 \ldots l_N}(t,\vec{p}-\vec{\Delta}/2,\vec{p}_2,\ldots,\vec{p}_N)\,. 
\end{align}
Using the flavor-basis $N$-particle wave function we find that, similarly to the single-particle Wigner function, the flavor-basis counterpart of the reduced single-particle Wigner function is given by $\varrho_{\alpha\beta}(t,\vec{x},\vec{p}) = U_{\alpha i}\, \varrho_{ij}(t,\vec{x},\vec{p})\, U^\dagger_{j\beta}$.

Substituting equation \eqref{eq:Npartwavefunc} into equation \eqref{eq:redwigfunc} and integrating over the momenta $\vec{p}_2$ to $\vec{p}_N$, we find that in the limit of non-overlapping individual wave functions the reduced single-particle Wigner function is related to the single-particle Wigner functions of the individual neutrinos by
\begin{align}
\label{eq:nonoverlaplimit}
\varrho_{ij}(t,\vec{x},\vec{p}) = \sum^N_{l=1} \varrho^{(l)}_{ij}(t,\vec{x},\vec{p})\,.
\end{align}
In the flavor basis we obtain a similar relation. Equation \eqref{eq:nonoverlaplimit} implies that in an experiment sensitive to the  charged-current interactions the count rate is proportional to 
\begin{align}
\label{eq:CCcountrate2}
\left. \frac{dn}{dt}\right|_{CC}  \propto
 \int_P d^3\vec{p} \int_V d^3\vec{x}\, \varrho_{ee}(t,\vec{x},\vec{p}) =  \int_P d^3\vec{p} \int_V d^3\vec{x}\,  \sum_l  \varrho^{(l)}_{ee}(t,\vec{x},\vec{p})\,.
\end{align}
Equations \eqref{eq:CCcountrate2} and \eqref{eq:CCcountrate1} differ only by the order of summation over the neutrino ensemble: while in equ\-ation \eqref{eq:CCcountrate1} the summation is performed at the very end, in equation  \eqref{eq:CCcountrate2} it is done in the beginning. Thus, the reduced single-particle Wigner function naturally reproduces the  prescription used in the single-particle quantum-mechanical approach to integrate over the neutrino production points and spectrum.  Similarly,  in an experiment detecting neutrinos via the neutral-current interactions the count rate is proportional to
\begin{align}
\label{eq:NCcountrate2}
\left. \frac{dn}{dt}\right|_{NC}  \propto
\sum_\alpha \int_P d^3\vec{p} \int_V d^3\vec{x}\, \varrho_{\alpha\alpha}(t,\vec{x},\vec{p}) =  \sum_\alpha \int_P d^3\vec{p} \int_V d^3\vec{x}\,  \sum_l  \varrho^{(l)}_{\alpha\alpha}(t,\vec{x},\vec{p})\,,
\end{align}
which reproduces equation \eqref{eq:NCcountrate1}. Equation \eqref{eq:nonoverlaplimit} together with equations \eqref{eq:NCcountrate2} and \eqref{eq:CCcountrate2} implies that the $N$-particle approach to neutrino oscillations simultaneously accounts for both sources of kinematic decoherence: dephasing of  different momentum modes of a single-neutrino wave-packet, related to the uncertainty principle, and dephasing of many neutrinos.

\paragraph{Evolution equation for the reduced single-particle Wigner function.}  
An evolution equation for the reduced single-particle Wigner function can be derived directly from the Schr\"odinger equation. Differentiating equation \eqref{eq:redwigfunc} with respect to time, subsequently using equation \eqref{eq:NparticleSchrodinger}, and taking into account hermiticity of the Hamiltonian we obtain 
\begin{align}
\label{eq:evoleqreducedWigner} 
\partial_t  \varrho_{ij}(t,\vec{x},\vec{p}) & =  -i \int \frac{d^3\vec{\Delta}}{(2\pi)^3} e^{i\vec{\Delta}\vec{x}}\, N \sum_{l_2\ldots l_N}\int \frac{d^3\vec{p}_2}{(2\pi)^3} \ldots \frac{d^3\vec{p}_N}{(2\pi)^3} \sum_{k_1\ldots k_N}
\int \frac{d^3\vec{k}_1}{(2\pi)^3} \ldots \frac{d^3\vec{k}_N}{(2\pi)^3}
\nonumber\\
& \times \bigl[{\sf H}_{il_2\ldots l_N\, k_1 \ldots k_N}(t,\vec{p}+\vec{\Delta}/2,\vec{p}_2,\ldots,\vec{p}_N,\vec{k}_1,\ldots,\vec{k}_N) \nonumber\\
&\times \psi_{k_1 \ldots k_N}(t,\vec{k}_1,\ldots,\vec{k}_N) \,  \psi^*_{j l_2\ldots l_N}(t,\vec{p}-\vec{\Delta}/2,\vec{p}_2,\ldots,\vec{p}_N)\nonumber\\
& - \psi_{il_2 \ldots l_N}(t,\vec{p}+\vec{\Delta}/2,\vec{p}_2,\ldots,\vec{p}_N)\,\psi^*_{k_1\ldots k_N}(t,\vec{k}_1,\ldots,\vec{k}_N)\nonumber\\
&\times   {\sf H}_{k_1\ldots k_N\, jl_2\ldots l_N}(t,\vec{k}_1,\ldots, \vec{k}_N,\vec{p}-\vec{\Delta}/2, \vec{p}_2,\ldots,\vec{p}_N) \bigr]\,.
\end{align}
Let us first consider those terms of the Hamiltonian \eqref{eq:NparticleHamiltonian} that contain ${\sf H}_{l_a k_a}(t,\vec{p}_a,\vec{k}_a)$ with $a > 1$, $\ldots + \delta_{ik_1}(2\pi)^3\delta(\vec{p}+\vec{\Delta}/2-\vec{k}_1) \ldots {\sf H}_{l_a k_a}(t,\vec{p}_a,\vec{k}_a)\ldots \delta_{l_Nk_N}(2\pi)^3\delta(\vec{p}_N-\vec{k}_N) + \ldots\,$. Their contribution to the right-hand side of  equation\eqref{eq:evoleqreducedWigner} reads,
\begin{align}
\label{eq:zerocontrib}
\partial_t \varrho_{ij}(t,\vec{x},\vec{p}) & \ni  -i \int \frac{d^3\vec{\Delta}}{(2\pi)^3} e^{i\vec{\Delta}\vec{x}} \,N\sum_{l_2\ldots l_N}\int \frac{d^3\vec{p}_2}{(2\pi)^3} \ldots \frac{d^3\vec{p}_a}{(2\pi)^3} \ldots  \frac{d^3\vec{p}_N}{(2\pi)^3} \sum_{k_a}
\int \frac{d^3\vec{k}_a}{(2\pi)^3}\nonumber\\
& \times \bigl[{\sf H}_{l_ak_a}(t,\vec{p}_a,\vec{k}_a) \psi_{il_2 \ldots k_a\ldots l_N}(t,\vec{p}+\vec{\Delta}/2,\vec{p}_2,\ldots,\vec{k}_a,\ldots,\vec{p}_N)\nonumber\\
& \cdot \psi^*_{jl_2\ldots l_a\ldots  l_N}(t,\vec{p}-\vec{\Delta}/2,\vec{p}_2,\ldots,\vec{p}_a,\ldots,\vec{p}_N)\nonumber\\
& - \psi_{il_2 \ldots l_a\ldots l_N}(t,\vec{p}+\vec{\Delta}/2,\vec{p}_2,\ldots,\vec{p}_a,\ldots,\vec{p}_N)\nonumber\\
& \cdot\psi^*_{jl_2\ldots k_a \ldots l_N}(t,\vec{p}-\vec{\Delta}/2,\vec{p}_2,\ldots,\vec{k}_a,\ldots,\vec{p}_N) {\sf H}_{k_al_a}(t,\vec{k}_a,\vec{p}_a)\bigr]\,.
\end{align}
Renaming $l_a \leftrightarrow k_a$ and $\vec{p}_l\leftrightarrow\vec{k}_l$ in the second term, we observe that  the right-hand side of equation \eqref{eq:zerocontrib} vanishes. That is, the ${\sf H}_{l_a k_a}(t,\vec{p}_a,\vec{k}_a)$ terms do not contribute to the evolution equation. The contribution of the remaining term is given by,
\begin{align}
\label{eq:nonzerocontrib}
\partial_t \varrho_{ij}(t,\vec{x},\vec{p}) &  \ni -i \int \frac{d^3\vec{\Delta}}{(2\pi)^3} e^{i\vec{\Delta}\vec{x}} \,N\sum_{l_2\ldots l_N}\int \frac{d^3\vec{p}_2}{(2\pi)^3}  \ldots  \frac{d^3\vec{p}_N}{(2\pi)^3} \sum_{k}
\int \frac{d^3\vec{k}}{(2\pi)^3}\nonumber\\
& \times \bigl[{\sf H}_{ik}(t,\vec{p}+\vec{\Delta}/2,\vec{k})\nonumber\\
& \times  \psi_{kl_2 \ldots l_N}(t,\vec{k},\vec{p}_2,\ldots,\vec{p}_N)\, \psi^*_{jl_2 \ldots l_N}(t,\vec{p}-\vec{\Delta}/2,\vec{p}_2,\ldots,\vec{p}_N)\nonumber\\
& - \psi_{il_2 \ldots l_N}(t,\vec{p}+\vec{\Delta}/2,\vec{p}_2,\ldots,	\vec{p}_N)\, \psi^*_{kl_2\ldots l_N}(t,\vec{k},\vec{p}_2,\ldots,\vec{p}_N)\nonumber\\
& \times 
{\sf H}_{kj}(t,\vec{k},\vec{p}-\vec{\Delta}/2)\bigr]\,,
\end{align}
where we have renamed $k_1 \rightarrow k$ and $\vec{k}_1 \rightarrow \vec{k}$ to emphasize its similarity to equation  \eqref{eq:timederrhoingom0}. Using equation \eqref{eq:twomomhamiltonian} we can recast it in the form,
\begin{align}
\label{eq:nonzerocontrib1}
\partial_t \varrho_{ij}(t,\vec{x},\vec{p}) &  \ni -i \int \frac{d^3\vec{\Delta}}{(2\pi)^3} e^{i\vec{\Delta}\vec{x}}\,N \sum_{l_2\ldots l_N}\int \frac{d^3\vec{p}_2}{(2\pi)^3}  \ldots  \frac{d^3\vec{p}_N}{(2\pi)^3} \sum_{k}
\int \frac{d^3\vec{k}}{(2\pi)^3}\int d^3\vec{y}\nonumber\\
& \times \bigl[e^{-i(\vec{p}+\vec{\Delta}/2-\vec{k})\vec{y}}{\sf H}_{ik}\bigl(t,\vec{y},{\textstyle \frac{\vec{p}+\vec{\Delta}/2+\vec{k}}{2}}\bigr)  \nonumber\\
&\times \psi_{kl_2 \ldots l_N}(t,\vec{k},\vec{p}_2,\ldots,\vec{p}_N) \psi^*_{jl_2\ldots l_N}(t,\vec{p}-\vec{\Delta}/2,\vec{p}_2,\ldots,\vec{p}_N)\nonumber\\
& - \psi_{il_2 \ldots n_N}(t,\vec{p}+\vec{\Delta}/2,\vec{p}_2,\ldots,\vec{p}_N)\psi^*_{kl_2\ldots l_N}(t,\vec{k},\vec{p}_2,\ldots,\vec{p}_N)\nonumber\\
& \times {\sf H}_{kj}\bigl(t,\vec{y},{\textstyle \frac{\vec{p}-\vec{\Delta}/2+\vec{k}}{2}}\bigr)e^{i(\vec{p}-\vec{\Delta}/2-\vec{k})\vec{y}} \bigr]\,.
\end{align}
Applying the same transformations that led from equation \eqref{eq:timederrhoingom} to equation \eqref{eq:timederrhoingom1} we arrive at 
\begin{align}
\label{eq:nonzerocontrib2}
\partial_t \varrho_{ij}(t,\vec{x},\vec{p}) &  \ni -i \int \frac{d^3\vec{\Delta}}{(2\pi)^3} e^{i\vec{\Delta}\vec{x}} \,N \sum_{l_2\ldots l_N}\int \frac{d^3\vec{p}_2}{(2\pi)^3}  \ldots  \frac{d^3\vec{p}_N}{(2\pi)^3} \sum_{k}
\int \frac{d^3\vec{k}}{(2\pi)^3}\int d^3\vec{y}e^{-i(\vec{p}-\vec{k})\vec{y}}\nonumber\\
& \times \bigl[{\sf H}_{ik}\bigr(t,\vec{x}+\vec{y}/2,\vec{p}+\vec{\Delta}/2\bigr) \nonumber\\
&\times \psi_{kl_2 \ldots l_N}(t,\vec{k}+\vec{\Delta}/2,\vec{p}_2,\ldots,\vec{p}_N) \psi^*_{jl_2 \ldots l_N}(t,\vec{k}-\vec{\Delta}/2,\vec{p}_2\ldots\vec{p}_N)\nonumber\\
& - \psi_{il_2 \ldots l_N}(t,\vec{k}+\vec{\Delta}/2,\vec{p}_2,\ldots\vec{p}_N)\psi^*_{kl_2\ldots l_N}(t,\vec{k}-\vec{\Delta}/2,\vec{p}_2,\ldots,\vec{p}_N)\nonumber\\
& \times {\sf H}_{kj}\bigl(t,\vec{x}-\vec{y}/2,\vec{p}-\vec{\Delta}/2\bigr)\bigr]\,.
\end{align}
Performing the same steps that led from equation \eqref{eq:timederrhoingom1} to equation \eqref{eq:timederrhoingom2}, we obtain an evolution equation for the reduced single-particle Wigner function. Its form matches the form of equation \eqref{eq:timederrhoingom2}. Performing the first-order gradient expansion,  we can recast the evolution equation in the familiar form,
\begin{align}
\label{eq:evoleqredWigner}
\partial_t\varrho(t,\vec{x},\vec{p}) &  + {\textstyle\frac12}\bigl\{\partial_\vec{p}{\sf H}(t,\vec{x},\vec{p}),\partial_\vec{x}\varrho(t,\vec{x},\vec{p})\bigr\} - {\textstyle\frac12}\bigl\{\partial_\vec{x}{\sf H}(t,\vec{x},\vec{p}),\partial_\vec{p}\varrho(t,\vec{x},\vec{p})\bigr\}\nonumber\\
& \approx -i \bigl[{\sf H}(t,\vec{x},\vec{p}),\varrho(t,\vec{x},\vec{p}) \bigr]\,.
\end{align}
Supplemented by the initial conditions constructed by combining equations \eqref{eq:Npartwavefunc} and \eqref{eq:redwigfunc}, equation \eqref{eq:evoleqredWigner} accounts for the Heisenberg uncertainty principle and the Pauli exclusion principle. This has long been known in the context of non-relativistic quantum-mechanical systems, see e.g. references \citep{2007cond.mat.3185C,PhysRevB.76.214301} and references therein. On the other hand, unlike the Schr\"odinger equation, the evolution equation also admits classical solutions describing an ensemble of particles with definite coordinates and momenta, if supplemented by the classical initial conditions, see appendix \ref{sec:classicallimit}.

The left-hand side of equation \eqref{eq:evoleqredWigner} is a continuity equation that conserves particle number. Its right-hand side causes only a unitarity rotation in the generation space and thus conserves ${\rm tr} \varrho(t,\vec{x},\vec{p})$. Therefore, if $\varrho(t_P,\vec{x},\vec{p})$ lies in the interval $[0,1]$ initially, as is required by the Pauli principle, then this trivially holds true at any $t\geq t_P$. The less trivial statement, and this is the main result of this section, is that quantum dynamics for the $N$-particle wave function actually leads to the evolution equation of the form \eqref{eq:evoleqredWigner}. Not only the contributions of the single-particle Wigner functions of the individual neutrinos, see equation \eqref{eq:nonoverlaplimit}, but also the `cross-terms' proportional to the overlap of the individual wave functions satisfy this evolution equation.
 	
\section{\label{sec:kinapproach}Kinetic approach}
The kinetic equation for the matrix of densities is derived in the framework of perturbative quantum field theory in the Heisenberg representation. Its form matches the form of the evolution equation for the (reduced) single-particle Wigner function. For a single neutrino system the initial matrix of densities is given by the single-particle Wigner function evaluated at $t=t_P$. For an $N$-particle neutrino system the initial matrix of densities is given by the reduced single-particle Wigner-function evaluated at $t=t_P$. Because the form of and the initial conditions for the evolution and kinetic equations match, their solutions are also identical. This implies in particular, that the uncertainty and exclusion principles can be accounted for also in the kinetic approach to oscillations  by considering initial conditions consistent with these fundamental quantum principles.

\paragraph{Kinetic equation for the matrix of densities.} 
In  astrophysical objects such as core-collapse supernovae or neutron-star mergers neutrino oscillations and collisions, including processes of neutrino emission and absorption, might be equally important. Processes changing the particle number are naturally accounted for in the second-quantized formalism. In a homogeneous system the particle number density is given by the expectation value of the operator $\hat{a}^\dagger_j(t,\vec{p})\hat{a}_i(t,\vec{p})$ constructed from the ladder operators of the neutrino field. In a system with weak inhomogeneities one should consider the expectation value of the operator 
\begin{align}
\label{eq:matrixofdensitiesoperator}
\hat{\varrho}_{ij}(t,\vec{x},\vec{p}) = \int \frac{d^3\vec{\Delta}}{(2\pi)^3}e^{i\vec{\Delta}\vec{x}} 
\hat{a}^\dagger_j(t,\vec{p}-\vec{\Delta}/2) \hat{a}_i(t,\vec{p}+\vec{\Delta}/2)\,
\end{align}
instead \citep{Sigl:1992fn,Stirner:2018ojk}. Let us emphasize that its arguments $\vec{x}$ and $\vec{p}$ are not coordinates and momenta of the field excitations. To keep this work self-contained, in this paragraph we repeat the derivation of the equation of motion for the operator $\hat{\varrho}_{ij}(t,\vec{x},\vec{p})$ presented in reference \citep{Stirner:2018ojk}. First, we differentiate equation \eqref{eq:matrixofdensitiesoperator} with respect to time. The time derivative of the ladder operators is determined by  the Heisenberg equation, e.g.
\begin{align}
\label{eq:heisenbergeq}
\partial_t \hat{a}_i(t,\vec{p}) = i\bigl[\hat{{\sf H}}, \hat{a}_i(t,\vec{p})\bigr]\,,
\end{align}
and similarly for $\hat{a}^\dagger(t,\vec{p})$. In the collisionless limit (but including forward scattering) we can write the Hamiltonian in the form \citep{Sigl:1992fn,Kartavtsev:2015eva,Stirner:2018ojk}
\begin{align}
\label{eq:bilinearhamiltonian}
\hat{\sf H} = \int \frac{d^3\vec{p}}{(2\pi)^3}\frac{d^3\vec{k}}{(2\pi)^3}\hat{a}^\dagger_i(t,\vec{p}){\sf H}_{ij}(t,\vec{p},\vec{k}) \hat{a}_i(t,\vec{k})\,,
\end{align}
with ${\sf H}_{ij}(t,\vec{p},\vec{k})$ defined in equation \eqref{eq:twomomhamiltonian}. Combining equations \eqref{eq:heisenbergeq} and \eqref{eq:bilinearhamiltonian} we arrive at
\begin{subequations}
\begin{align}
\partial_t \hat{a}_i(t,\vec{p}) & = -\,i \int\frac{d^3\vec{k}}{(2\pi)^3} {\sf H}_{in}(t,\vec{p},\vec{k}) \hat{a}_n(t,\vec{k})\,,\\
\partial_t \hat{a}^\dagger_j(t,\vec{p}) & = +\, i\int\frac{d^3\vec{k}}{(2\pi)^3} \hat{a}^\dagger_n(t,\vec{k}) {\sf H}_{nj}(t,\vec{k},\vec{p}) \,.
\end{align}
\end{subequations}
The resulting expression for the time derivative of $\hat{\varrho}$ reads  
\begin{align}
\label{eq:eqforrhohat}
\partial_t\hat{\varrho}_{ij}(t,\vec{x},\vec{p}) & = -i \int \frac{d^3\vec{\Delta}}{(2\pi)^3}e^{i\vec{\Delta}\vec{x}} \int\frac{d^3\vec{k}}{(2\pi)^3}\nonumber\\
& \times [{\sf H}_{in}(t,\vec{p}+\vec{\Delta}/2,\vec{k}) \, \hat{a}^\dagger_j(t,\vec{p}-\vec{\Delta}/2) \hat{a}_n(t,\vec{k})\nonumber\\
&-\hat{a}^\dagger_n(t,\vec{k}) \hat{a}_i(t,\vec{p}+\vec{\Delta}/2)\,  {\sf H}_{nj}(t,\vec{k},\vec{p}-\vec{\Delta}/2)]\,.
\end{align}
Using equation \eqref{eq:twomomhamiltonian} we can rewrite the equation of motion in the form  
\begin{align}
\partial_t\hat{\varrho}_{ij}(t,\vec{x},\vec{p}) & = -i \int \frac{d^3\vec{\Delta}}{(2\pi)^3}e^{i\vec{\Delta}\vec{x}} \int\frac{d^3\vec{k}}{(2\pi)^3}\int d^3\vec{y}\nonumber\\
& \times \bigl[e^{-i(\vec{p}+\vec{\Delta}/2-\vec{k})\vec{y}} {\sf H}_{ik}\bigl(t,\vec{y},{\textstyle\frac{\vec{p}+\vec{\Delta}/2+\vec{k}}{2}\bigr) \, \hat{a}^\dagger_j(t,\vec{p}-\vec{\Delta}/2)}\hat{a}_k(t,\vec{k}) \nonumber\\
&-\hat{a}^\dagger_k(t,\vec{k}) \hat{a}_i(t,\vec{p}+\vec{\Delta}/2)\, {\sf H}_{kj}\bigl(t,\vec{y},{\textstyle \frac{\vec{p}-\vec{\Delta}/2+\vec{k}}{2}}\bigr)e^{i(\vec{p}-\vec{\Delta}/2-\vec{k})\vec{y}} \bigr]\,.
\end{align}
Using the same transformations that led from equation \eqref{eq:timederrhoingom} to equation \eqref{eq:timederrhoingom1} we find
\begin{align}
\partial_t\hat{\varrho}_{ij}(t,\vec{x},\vec{p}) & = -i \int \frac{d^3\vec{\Delta}}{(2\pi)^3} e^{i\vec{\Delta}\vec{x}} \int \frac{d^3\vec{k}}{(2\pi)^3} \int d^3 \vec{y} e^{-i(\vec{p}-\vec{k})\vec{y}}\nonumber\\
& \times \bigl[\mathsf{H}_{ik}\bigl(t,\vec{x}+\vec{y}/2, \vec{p}+\vec{\Delta}/2\bigr)\hat{a}^\dagger_j(t,\vec{k}-\vec{\Delta}/2) \hat{a}_k(t,\vec{k}+\vec{\Delta}/2)\nonumber\\
& - \hat{a}^\dagger_k(t,\vec{k}-\vec{\Delta}/2)  \hat{a}_i(t,\vec{k}+\vec{\Delta}/2)\mathsf{H}_{kj}(t,\vec{x}-\vec{y}/2,\vec{p}-\vec{\Delta}/2\bigr)\bigr]\,.
\end{align}
Expanding the Hamiltonian and carrying out the integrations we obtain a quantum-field-theoretical counterpart of equation \eqref{eq:timederrhoingom2}, 
\begin{align}
\label{eq:eqofmotion}
\partial_t\hat{\varrho}_{ij}(t,\vec{x},\vec{p}) & =-i\sum_{n,m} \frac{(-i)^{n-m}}{2^{n+m}\, n!\,m!} \bigl[
\partial^m_\vec{x}\partial^n_\vec{p}\mathsf{H}_{ik}(t,\vec{x},\vec{p})\partial^m_\vec{p}\partial^n_\vec{x}\hat{\varrho}_{kj}(t,\vec{x},\vec{p}) \nonumber\\
&-(-1)^{n+m}\partial^m_\vec{p}\partial^n_\vec{x}\hat{\varrho}_{ik}(t,\vec{x},\vec{p})\partial^m_\vec{x}\partial^n_\vec{p}\mathsf{H}_{kj}(t,\vec{x},\vec{p})
\bigr]\,.
\end{align}
Next we take the expectation value of equation \eqref{eq:eqofmotion}, which amounts to the substitution $\hat{\varrho}\! \rightarrow\! \varrho\! \equiv\!\langle \hat{\varrho} \rangle$. Performing the first-order gradient expansion of the resulting kinetic equation for the matrix of densities we arrive at \citep{Sigl:1992fn,Stirner:2018ojk} 
\begin{align}
\label{eq:kineticequation}
\partial_t\varrho(t,\vec{x},\vec{p}) & + {\textstyle \frac12}\{\partial_\vec{p} \mathsf{H}(t,\vec{x},\vec{p}),\partial_\vec{x}\varrho(t,\vec{x},\vec{p})\} - {\textstyle \frac12}\{\partial_\vec{x} \mathsf{H}(t,\vec{x},\vec{p}),\partial_\vec{p}\varrho(t,\vec{x},\vec{p})\}\nonumber\\
& = -i [ \mathsf{H}(t,\vec{x},\vec{p}),\varrho(t,\vec{x},\vec{p})]\,.
\end{align}
As has been emphasized in reference \cite{Stirner:2018ojk}, in the relativistic limit the first anticommutator on its left-hand contains the full flavor-dependent velocity structure of the Liouville term. 

In dense environments neutrino-neutrino forward scattering, combined with flavor oscillations, generates flavor-off-diagonal components of the potential \citep{Pantaleone:1992eq}. In supernovae the as\-sociated energy scale, $\mu \sim \sqrt{2}G_F\,n_\nu$, is of the order of $10^{-5}$~eV. The resulting collective neutrino oscillation have been extensively studied  using equation \eqref{eq:kineticequation}, see e.g. references  \cite{Duan:2010bg,Mirizzi:2015eza,Chakraborty:2016yeg,Horiuchi:2017sku} and re\-fe\-rences therein. The recent works \citep{Tamborra:2017ubu,Abbar:2018shq,Azari:2019jvr,Morinaga:2019wsv,DelfanAzari:2019tez,Abbar:2019zoq,Nagakura:2019sig,Johns:2019izj,Glas:2019ijo,Capozzi:2019lso,Chakraborty:2019wxe,Shalgar:2019qwg,Capozzi:2020kge,shalgar2020dispelling,Padilla-Gay:2020uxa,Shalgar:2020xns} focus primarily on the so-called fast neutrino flavor conversions put forward in references \citep{Sawyer:2005jk,Sawyer:2008zs}. As has been argued in reference \citep{Capozzi:2018clo}, the fast flavor conversions can be triggered by the collision term, absent in equation \eqref{eq:kineticequation}. The collision term has been derived in reference \citep{Sigl:1992fn} using perturbative quantum field theory. In references \citep{Yamada:2000za,Vlasenko:2013fja,Cirigliano:2014aoa,Vlasenko:2014bva,Blaschke:2016xxt} the kinetic equation and the collision term have been derived using methods of non-equilibrium quantum field theory. The quantum-kinetic equation has been used in reference \citep{Richers:2019grc} to study flavor decoherence caused by irreversible entanglement with the environment.
 
\paragraph{Single-particle initial conditions.}
At the production time $t=t_P$ the state vector in the Schr\"odinger picture, used in the single- and $N$-particle quantum-mechanical approaches, coincides with the state vector in the Heisenberg picture, used in the kinetic approach. This allows us to construct initial conditions for the matrix of densities corresponding to those for the neutrino wave function. 

Equation \eqref{eq:eqofmotion} is valid for arbitrary initial conditions, including the single-particle ones. The state vector of the $i$'th mass eigenstate of momentum $\vec{p}$ reads $\hat{a}^\dagger_i(t_P,\vec{p})|0\rangle$. The respective wave packet is constructed as $|\psi_i\rangle = \int d^3\vec{p}/(2\pi)^3 \psi_i(t_P,\vec{p})\hat{a}^\dagger_i(t_P,\vec{p})|0\rangle$, where $\psi_i(t_P,\vec{p})$ is the $i$'th component of the neutrino wave function. Using orthogonality of the state vectors of different mass eigenstates, $\langle 0| \hat{a}_j(t_P,\vec{k})\hat{a}^\dagger_i(t_P,\vec{q})|0\rangle = (2\pi)^3\delta(\vec{k}-\vec{q})\,\delta_{ij}$, we obtain the following expression for the one-particle state vector in the Heisenberg picture `initiated' at $t_P$,
\begin{align}
\label{eq:onepartstatevector}
|\psi\rangle = \sum_i \int\frac{d^3\vec{p}}{(2\pi)^3} \psi_i(t_P,\vec{p}) \hat{a}^\dagger_i(t_P,\vec{p})|0\rangle\,.
\end{align}
Taking the expectation value of equation \eqref{eq:matrixofdensitiesoperator}  with respect to the state vector equation \eqref{eq:onepartstatevector} we obtain an expression for the matrix of densities identical to the  single-particle Wigner function,
\begin{align}
\varrho_{ij}(t_P,\vec{x},\vec{p}) = \int \frac{d^3\vec{\Delta}}{(2\pi)^3}e^{i\vec{\Delta}\vec{x}}\psi_i(t_P,\vec{p}+\vec{\Delta}/2)\psi^*_j(t_P,\vec{p}-\vec{\Delta}/2)\,.
\end{align}
In other words, the constructed initial conditions for the kinetic equation match those for the evolution equation for the single-particle Wigner function. Given that the two equations have the same form they produce identical solutions.  Hence, solutions of the kinetic equation account for the Heisenberg uncertainty principle if the initial conditions are consistent with this fundamental quantum principle.

To provide an explicit example let us return to the Gaussian wave packet specified in equation \eqref{eq:gaussianwp}. The resulting initial matrix of densities reads in the flavor basis
\begin{align}
\label{eq:singlepartmatrofdens}
\varrho_{\alpha\beta}(t_P,\vec{x},\vec{p}) = \delta_{\alpha e}\delta_{e\beta}\cdot 2^3\exp\biggl(-\frac{(\vec{p}-\vec{p}_w)^2}{2\sigma_p^2}\biggr)\,
\exp\biggl(-\frac{(\vec{x}-\vec{x}_P)^2}{2\sigma^2_x}\biggr)\,.
\end{align}
Note that $\varrho_{\alpha\beta}(t_P,\vec{x},\vec{p})$ is localized neither in coordinate nor in momentum space, thereby reflecting the Heisenberg uncertainty principle. The dephasing of its momentum modes in the course of the neutrino propagation is the source of kinematic decoherence for single neutrinos. The authors of references \citep{Kersten:2015kio, Akhmedov:2017mcc} argued that, due to very high temperatures and densities, in supernovae the neutrino wave packet is very small, $\sigma_x \sim 10^{-11}$~cm,  which corresponds to $\sigma_p\sim 1$~MeV. Note that in this case $\sigma_p$ is comparable to the neutrino momentum and the on-shell approximation used to derive the kinetic equation is likely to break down.
 
\paragraph{$N$-particle initial conditions.}
The kinetic equation has originally \citep{Sigl:1992fn} been formulated to study flavor neutrino evolution of multiparticle systems in the mean-field approximation. The state vector of an $N$-particle neutrino system is constructed as 
\begin{align}
\label{eq:Npartstatevector}
|\psi\rangle &= \frac{1}{\sqrt{N!}} \sum_{i_1\ldots i_N}\int \frac{d^3\vec{p}_1}{(2\pi)^3}\ldots \frac{d^3\vec{p}_N}{(2\pi)^3}\nonumber\\
&\times \psi_{i_1\ldots i_N}(t_P,\vec{p}_1\ldots \vec{p}_N)\,
\hat{a}^\dagger_{i_1}(t_P,\vec{p}_1) \ldots \hat{a}^\dagger_{i_N}(t_P,\vec{p}_N) |0\rangle\,,
\end{align}
where $\psi_{i_1\ldots i_N}(t_P,\vec{p}_1\ldots \vec{p}_N)$ is the $N$-particle wave function. For a system of independent neutrinos it is given by equation \eqref{eq:Npartwavefunc}. The wave function of a system with strong interactions does not factorize into a product of the individual wave functions. In either case the Pauli principle requires the wave function to change sign under permutations of any two particles. For fermions the ladder operators anticommute and $\hat{a}^\dagger_{i_1}(t_P,\vec{p}_1) \ldots \hat{a}^\dagger_{i_N}(t_P,\vec{p}_N) |0\rangle$ changes its sign as well. Therefore, the state vector remains invariant. Taking the expectation value of equation \eqref{eq:matrixofdensitiesoperator} with respect to  equation \eqref{eq:Npartstatevector}, we obtain an expression for the matrix of densities identical to the reduced single- particle Wigner function,
\begin{align}
\varrho_{ij}(t_P,\vec{x},\vec{p}) & = N\int \frac{d^3\vec{\Delta}}{(2\pi)^3}e^{i\vec{\Delta}\vec{x}}\sum_{l_2\ldots l_N}
\int \frac{d\vec{p}_2}{(2\pi)^3}\ldots \frac{d^3\vec{p}_N}{(2\pi)^3}\nonumber\\
& \times \psi_{i l_2\ldots l_N}(\vec{p}+\vec{\Delta}/2,\vec{p}_2,\ldots,\vec{p}_N)
\psi^*_{j l_2\ldots l_N}(\vec{p}-\vec{\Delta}/2,\vec{p}_2,\ldots,\vec{p}_N)\,.
\end{align}
In other words, the constructed initial conditions for the kinetic equation match those for the evolution equation for the reduced single-particle Wigner function. Given that the two equations have the same form they produce identical solutions. Hence, solutions of the kinetic equation account for the Heisenberg uncertainty principle as well as the Pauli exclusion principle if the initial conditions are consistent with these fundamental quantum principles.

To provide an explicit example, let us consider the wave function specified in equation \eqref{eq:Npartwavefunc} in the limit of non-overlapping individual wave functions. The exchange terms become sizeable only at extremely high densities which are far above those present in supernovae \cite{Akhmedov:2017mcc}, which justifies this approximation. The resulting initial matrix of densities is given by 
\begin{align}
\label{eq:matrofdensexample}
\varrho_{ij}(t_P,\vec{x},\vec{p}) = \sum^N_{l=1} \varrho^{(l)}_{ij}(t_P,\vec{x},\vec{p})\,,
\end{align} 
where \smash{$\varrho^{(l)}_{ij}(t_P,\vec{x},\vec{p})$} refer to the individual neutrinos, compare with equation \eqref{eq:nonoverlaplimit}. For Gaussian wave packets \smash{$\varrho^{(l)}_{ij}(t_P,\vec{x},\vec{p})$} are given by equation \eqref{eq:singlepartmatrofdens} with (in general) different $\vec{p}_w$ and $\vec{x}_P$ for different $l$. For a small $\sigma_p$ the spectrum of the neutrino ensemble is determined by the distribution of the characteristic momentum $\vec{p}_w$. On the other hand, a large $\sigma_p$ ``superimposed'' on the  distribution of $\vec{p}_w$ may leave an imprint on the observed neutrino spectrum. 

Equation \eqref{eq:matrofdensexample} implies that the kinetic equation is capable of accounting for both sources of kine\-mat\-ic decoherence: dephasing of  different momentum modes of a single-neutrino wave-packet, related to the uncertainty principle, and dephasing of many neutrinos. Their influence on collective neutrino oscillations has been addressed in references \citep{Akhmedov:2016gzx} and \citep{Raffelt:2010za} respectively. The authors of reference \citep{Akhmedov:2016gzx} argued that collective neutrino oscillations can only take place if the product of the neutrino-neutrino interaction potential $\mu$ and the initial length $P_0$  of the global flavor ``spin'' vector exceeds the neutrino momentum uncertainty $\sigma_p$. Therefore, the existence of the additional energy scale, related to the Heisenberg principle, may have an impact on the phenomenology of supernovae neutrinos via the effect of wave packet separation. 

Let us note that in the kinetic approach it is relatively straightforward to consider more complex initial states, for example the Glauber state addressed in reference \citep{PhysRevD.45.1782} or states with nonzero particle-antiparticle correlations addressed in references \citep{Volpe:2013uxl,Vaananen:2013qja,Serreau:2014cfa,Volpe:2015rla,Kartavtsev:2015eva}. In the latter case the kinetic equation for the matrix of densities is supplemented by evolution equations for the pair correlators.

\section{\label{sec:conclusions}Summary and conclusion}
The present work shows that for collisionless neutrino propagation (in a background medium) the quantum-mechanical and kinetic approaches to neutrino oscillations produce equivalent results. On the one hand, this equivalence means that all the results obtained so far in the quantum-mechanical approach can be reproduced in the kinetic approach. On the other hand, it means that many aspects of the quantum-mechanical approach can be directly transferred to the kinetic approach. Based on this observation, we also argue that solutions of the kinetic equation account for the Heisenberg uncertainty principle and the related effect of wave packet separation (for single neutrinos), as well as the Pauli exclusion principle (for $N$-particle systems) if the initial conditions for the matrix of densities are consistent with these fundamental quantum principles. This implies  that the neutrino momentum uncertainty is an integral part of the initial conditions for the matrix of densities.

To this end, we consider single- and $N$-particle neutrino systems and `translate' the Schr\"odinger equation for their wave functions, $\psi_i(t,\vec{p})$ and $\psi_{i_1\ldots i_N}(t,\vec{p}_1,\ldots,\vec{p}_N)$, into an evolution equation for the (reduced) single-particle Wigner function $\varrho_{ij}(t,\vec{x},\vec{p})$. The term `reduced' refers to the $N$-particle systems and means that coordinates and momenta of $N-1$ particles are integrated over. Both the single-particle and the reduced single-particle  Wigner function satisfy the same evolution equation. Solutions of the evolution equation, supplemented by initial conditions constructed from the neutrino wave function, at any given time $t$ reproduce the Wigner function calculated using $\psi_i(t,\vec{p})$ or $\psi_{i_1\ldots i_N}(t,\vec{p}_1,\ldots,\vec{p}_N)$ respectively. Because the wave function accounts for the Heisenberg uncertainty principle and, for $N$-particle systems, also for the Pauli exclusion principle, this applies also to the (reduced) single-particle Wigner function. This implies that the evolution equation for the neutrino Wigner function is capable of reproducing the standard results for the day-time and night-time neutrinos, including the impact of wave packet separation on the neutrino flavor composition. Because (in the collisionless limit) the form of the kinetic equation for the matrix of densities matches the form of the evolution equation for the Wigner function, this applies also to the kinetic equation, if the same initial conditions are used in the latter. Constructing the neutrino state vector using the wave function of the single- or $N$-particle system, we find that this requirement is automatically fulfilled.

The kinetic equation is a powerful tool for analysis of supernovae neutrinos, capable of descri\-bing neutrino propagation from the neutrinosphere, where collisions dominate, to the outer layers, dominated by oscillations. The  Pauli-blocking factors in its collision term ensure that the exclusion principle is fulfilled in the scattering processes. As shown here, its Liouville and oscillation terms  ensure that the matrix of densities remains consistent with the Pauli and Heisenberg principles if it was initially consistent with these principles. The Liouville and oscillation terms are responsible for both sources of kinematic decoherence: dephasing of many neutrinos, and dephasing of different momentum modes of a single-neutrino wave-packet. While the former is related to the energy spectrum characterizing the entire neutrino ensemble, the latter is related to the additional energy scale characterizing the size of the neutrino wave packets and thus to the uncertainty principle. The existence of this additional energy scale affects collective neutrino oscillations through the effect of kinematic decoherence. Further analysis of the interplay between the many energy scales involved -- the spectrum, the vacuum oscillation frequency, the matter potential, the neutrino-neutrino refraction potential, and the  neutrino momentum uncertainty -- may unravel new facets of the already very rich and complex phenomenology of neutrino oscillations in core collapse supernovae. An important prerequisite for this analysis is a detailed study of the shape and size of the neutrino wave packets in supernovae, for which only relatively crude estimates exist at present.

%%%%%%%%%%%%%%%%%%%%%%%%%%%%%%%%%%%%%%%%%%%%%
\acknowledgments
We acknowledge support by the Russian Science Foundation under the Grant No.18-72-10070.  The author would like to thank Alexandra Dobrynina and Georg G. Raffelt for valuable comments and careful proof-reading of the manuscript.
%%%%%%%%%%%%%%%%%%%%%%%%%%%%%%%%%%%%%%%%%%%%%

%%%%%%%%%%%%%%%%%%%%%%%%%%%%%%%%%%%%%%%%%%%%%
\appendix
%%%%%%%%%%%%%%%%%%%%%%%%%%%%%%%%%%%%%%%%%%%%%

\section{\label{sec:apprthreeflavor}Approximate three-flavor evolution}

In the flavor basis the Hamiltonian of a neutrino propagating in the background matter of the Sun or the Earth reads
\begin{align}
{\sf H} = U(\theta_{23},\theta_{13},\theta_{12},\delta)\, {\sf \Omega}\, U^\dagger(\theta_{23},\theta_{13},\theta_{12},\delta) + {\sf V}\,,
\end{align}
where $\mathsf{\Omega} = \mathrm{diag}(\omega_1,\omega_2,\omega_3)$ is the diagonal in the mass basis matrix of the kinetic energies, and ${\sf V}= {\rm diag}({\sf V}_e,0,0)$ is the diagonal in the flavor basis matrix of the matter potentials. In the standard parametrization the Pontecorvo-Maki-Nakagawa-Sakata (PMNS) matrix reads,
\begin{align}
\label{eq:UPMNS}
U(\theta_{23},\theta_{13},\theta_{12},\delta)&=O_{23}(\theta_{23})\Gamma_\delta O_{13}(\theta_{13}) \Gamma^\dagger_\delta O_{12}(\theta_{12})\,,
\end{align}
where $O_{ij}$ is the orthogonal rotation matrix in the $ij$-plane which depends on the mixing angle $\theta_{ij}$, and $\Gamma_\delta = {\rm diag}(1, 1, e^{\delta})$, with  $\delta$ being the Dirac-type CP-violating phase \citep{PhysRevD.98.030001}.

Following reference \citep{Akhmedov:2004rq} we consider another basis, ${\sf H} \rightarrow U^\dagger(\theta_{23},\theta_{13},\delta)\, {\sf H}\, U(\theta_{23},\theta_{13},\delta)$, where $U(\theta_{23},\theta_{13},\delta) \equiv O_{23}(\theta_{23})\Gamma_\delta O_{13}(\theta_{13})$. In this basis the Hamiltonian takes the form 
\begin{align}
{\sf H} \rightarrow \omega_1\cdot\mathbb{1} + 
\begin{pmatrix}
\Delta\omega_{21}s^2_{12}+ {\sf V}_e c^2_{13} & \Delta\omega_{21}s_{12}c_{12} & {\sf V}_e s_{13} c_{13} \\
\Delta\omega_{21} s_{12}c_{12} & \Delta\omega_{21} c^2_{12} & 0 \\
{\sf V}_e s_{13}c_{13} & 0 & \Delta\omega_{31}+{\sf V}_e s^2_{13}
\end{pmatrix}\,,
\end{align} 
where $\mathbb{1}$ is the identity matrix. The approximation scheme developed in reference \citep{Akhmedov:2004rq} consists in neglecting the $(1,3)$ and $(3,1)$ elements of this matrix, i.e. the terms 
\begin{align}
\Delta {\sf H} = {\sf V}_e s_{13}c_{13} 
\begin{pmatrix}
0 & 0 & 1 \\ 
0 & 0 & 0 \\
1 & 0 & 0
\end{pmatrix}\,.
\end{align}
Let us note that the authors of reference \citep{Ioannisian:2018qwl} developed an approximation scheme that takes into account also the $\Delta\mathsf{H}$ terms. Rotating $\Delta \mathsf{H}$ to the mass basis we obtain 
\begin{align}
\Delta{\sf H} \rightarrow O^\dagger_{12}(\theta_{12})\Gamma_\delta\,\Delta {\sf H}\, \Gamma^\dagger_\delta O_{12}(\theta_{12}) = {\sf V}_e s_{13}c_{13}
\begin{pmatrix}
0 & 0 &  c_{12}e^{-i\delta} \\
0 & 0 &  s_{12}e^{-i\delta} \\
c_{12}e^{i\delta} & s_{12}e^{i\delta} & 0
\end{pmatrix}\,.
\end{align}
Subtracting these terms from the Hamiltonian in the mass basis,
\begin{align}
{\sf H} = {\sf \Omega} + U^\dagger(\theta_{23},\theta_{13},\theta_{12},\delta)\,{\sf V}\,U(\theta_{23},\theta_{13},\theta_{12},\delta)\,,
\end{align}
we arrive at the mass-basis counterpart of the approximate Hamiltonian studied in reference \citep{Akhmedov:2004rq},
\begin{align}
\label{eq:Happox}
{\sf H}(t,\vec{p}) \approx 
\begin{pmatrix}
\omega_1(\vec{p})+{\sf V}_e(t)\,c^2_{13} c^2_{12} & {\sf V}_e(t)\,c^2_{13} c_{12}s_{12} & 0 \\ 
{\sf V}_e(t)\,c^2_{13} c_{12}s_{12}  & \omega_2(\vec{p}) + {\sf V}_e(t)\,c^2_{13} s^2_{12}  & 0\\
0  & 0 & \omega_3(\vec{p})+{\sf V}_e(t)\,s^2_{13}
\end{pmatrix} \,.
\end{align}
Equation \eqref{eq:Happox} is diagonalized by a rotation in the 12-plane, $O_{12}^\dagger(\vartheta_{12}){\sf H} O_{12}(\vartheta_{12})  = {\sf E}$, with the rotation angle given by 
\begin{align}
\tan 2\vartheta_{12}(t,\vec{p}) = \frac{{\sf V}_e(t) c^2_{13}\cdot\sin{2\theta_{12}}}{\Delta\omega_{21}(\vec{p})-{\sf V}_e(t) c^2_{13}\cdot\cos{2\theta_{12}}}\,,
\end{align}
where $\Delta\omega_{ij}(\vec{p}) \equiv \omega_i(\vec{p})-\omega_j(\vec{p})$. For $V_e = 0$  the Hamiltonian is diagonal and therefore $\vartheta_{12} = 0$. For ${\sf V}_e(t) c^2_{13}\cdot\cos{2\theta_{12}}\rightarrow \Delta\omega_{21}(\vec{p})$ the resonance occurs, $\tan 2\vartheta_{12}\rightarrow\infty$, and therefore  $\vartheta_{12}\rightarrow \pi/4$. In the limit ${\sf V}_e(t) c^2_{13}\cdot\cos{2\theta_{12}} \gg \Delta\omega_{21}(\vec{p})$ we find $\tan 2\vartheta_{12} \rightarrow -\tan 2\theta_{12}$. Requiring the continuity of $\vartheta_{12}$ (considered as a function of $\mathsf{V}_e$) we obtain $\vartheta_{12} \rightarrow \pi/2 - \theta_{12}$ in this case.  Upon the diagonalization we obtain  $\mathsf{E}=diag (\bar{E}-\Delta E/2,\bar{E}+\Delta E/2,E_3)$ with
\begin{subequations}
\label{eq:Eeigenvals}
\begin{align}
\bar{E}& = \bigl[\omega_1(\vec{p}) + \omega_2(\vec{p})+{\sf V}_e(t) c^2_{13}\bigr]/2 \,,\\
\Delta E & = \bigl[(\Delta \omega_{21}(\vec{p})-{\sf V}_e(t) c^2_{13} \cdot \cos 2\theta_{12})^2 + ({\sf V}_e(t) c^2_{13} \cdot \sin 2\theta_{12})^2\bigr]^\frac12\,.
\end{align}
\end{subequations}
Although written in a slightly different form, equation \eqref{eq:Eeigenvals} is identical to the one obtained in the flavor basis  \citep{deHolanda:2003nj,Akhmedov:2017mcc}. Note that $\Delta E$ (considered as a function of $\vec{p}$) reaches its minimum at the resonance. 

\section{\label{sec:adiabaticprop}Time evolution operator in the adiabatic limit}
The time evolution operator satisfies a Schr\"odinger-like equation,
\begin{align}
\label{eq:evoleqU}
\partial_t {\sf U}(t,t_P,\vec{p}) = -i {\sf H}(t,\vec{p}) {\sf U}(t,t_P,\vec{p}) \,.
\end{align}
As has been discussed in appendix \ref{sec:apprthreeflavor}, in the approximation  used here the Hamiltonian is diagonalized by a rotation in the 12-plane, $\mathsf{E}(t,\vec{p}) = O^\dagger_{12}(\vartheta_{12}(t,\vec{p}))\mathsf{H}(t,\vec{p})O_{12}(\vartheta_{12}(t,\vec{p}))$. Using equation \eqref{eq:evoleqU} we obtain the following evolution equation for the matrix $O^\dagger_{12}(\vartheta_{12}(t,\vec{p})){\sf U}(t,t_P,\vec{p})$:
\begin{align}
\partial_t[O^\dagger_{12}(\vartheta_{12}(t,\vec{p}))&\mathsf{U}(t,t_P,\vec{p})]  = -i\mathsf{E}(t,\vec{p}) [\,O^\dagger_{12}(\vartheta_{12}(t,\vec{p}))\mathsf{U}(t,t_P,\vec{p})] \nonumber \\ 
& +[\partial_t O^\dagger_{12}(\vartheta_{12}(t,\vec{p}))\cdot O_{12}(\vartheta_{12}(t,\vec{p}))][O^\dagger_{12}(\vartheta_{12}(t,\vec{p}))\mathsf{U}(t,t_P,\vec{p})]\,.
\end{align}
The term $\partial_t O^\dagger_{12}(\vartheta_{12}(t,\vec{p}))\cdot O_{12}(\vartheta_{12}(t,\vec{p}))\propto \partial_t \vartheta_{12}(t,\vec{p})$ can be neglected in the adiabatic limit. Solution of the resulting approximate equation reads
\begin{align}
O^\dagger_{12}(\vartheta_{12}(t,\vec{p}))\mathsf{U}(t,t_P,\vec{p}) = \exp\left(-i\int^t_{t_P} \mathsf{E}(\tau,\vec{p})d\tau\right)
O^\dagger_{12}(\vartheta_{12}(t_P,\vec{p}))\mathsf{U}(t_P,t_P,\vec{p})\,.
\end{align}
Taking into account the initial conditions, ${\sf U}(t_P,t_P,\vec{p}) = \mathbb{1}$, we finally arrive at
\begin{align}
\label{eq:evolopadiabatic}
\mathsf{U}(t,t_P,\vec{p}) = O_{12}(\vartheta_{12}(t,\vec{p}))\exp\left(-i\int^t_{t_P} \mathsf{E}(\tau,\vec{p})d\tau\right)
O^\dagger_{12}(\vartheta_{12}(t_P,\vec{p}))\,.
\end{align}

\section{\label{sec:kindecoherday}Propagation in the Sun}
Applying equation \eqref{eq:evolopadiabatic} to the neutrino propagation in the Sun we obtain 
\begin{align}
{\sf U}(t_S,t_P,\vec{p}) = O_{12}(\vartheta_{12}(t_S,\vec{p})) \exp\left(-i\int^{t_S}_{t_P} \mathsf{E}(\tau,\vec{p})d\tau\right) O^\dagger_{12}(\vartheta_{12}(t_P,\vec{p}))\,,
\end{align}
where components of the diagonal matrix $E$ can be read off from equations \eqref{eq:Happox} and \eqref{eq:Eeigenvals}.  As an overall phase cancels out in the Wigner function, it is convenient to subtract $\int^{t_S}_{t_P} \bar{E}(\tau,\vec{p})d\tau$ from the integral. The $(1,1)$ and $(2,2)$ terms of the phase exponent are then determined by $\Delta E$. Hence, $\Delta\phi_{12}(t_S,t,_P,\vec{p}) = -\int^{t_S}_{t_P} \Delta E(\tau,\vec{p})d\tau$ and $\partial_\vec{p}\Delta\phi_{12}(t_S,t,_P,\vec{p}) = -\int^{t_S}_{t_P} \partial_\vec{p}\Delta E(\tau,\vec{p})d\tau$. Using equation \eqref{eq:Eeigenvals} we obtain \citep{MIKHEYEV198941,deHolanda:2003nj,Akhmedov:2017mcc}
\begin{align}
\partial_\vec{p}\Delta\phi_{12}(t_S,t,_P,\vec{p}) = -\int^{t_S}_{t_P} \!d\tau\, \frac{\Delta \omega_{21}(\vec{p})-\mathsf{V}_ec^2_{13}\cdot \cos{2\theta_{12}}}{(\Delta \omega_{21}(\vec{p})-{\sf V}_e(t) c^2_{13} \cdot \cos 2\theta_{12})^2 + ({\sf V}_e(t) c^2_{13} \cdot \sin 2\theta_{12})^2}\,.
\end{align}
The integrand vanishes at the MSW resonance and changes its sign after crossing the resonance.  As has been emphasized in reference \citep{deHolanda:2003nj}, this effect can partially alleviate the build-up of kinematical decoherence for neutrinos produced above the resonance.  

\section{\label{sec:nightneutrinos}Day-night asymmetry}

Due to the spatial separation of the neutrino wave packet during its propagation from the production point to the detector, $|{\sf U}_{ee}(t_N,t_P,\vec{p})|^2$  can be approximated by  
\begin{align}
|{\sf U}_{ee}(t_N,t_P,\vec{p})|^2 = \sum_i |[U(\theta_{23},\theta_{13},\theta_{12},\delta) {\sf U}(t_N,t_D,\vec{p})]_{ei}|^2|U^\dagger_{ie}(\theta_{23},\theta_{13},\theta^P_{12},\delta)|^2 \,.
\end{align}
The Earth matter effects are small and hence ${\sf U}(t_N,t_D,\vec{p})$ is close to $\mathbb{1}$. Following reference \citep{Goswami:2004cn} we parametrize a deviation of $|[U(\theta_{23},\theta_{13},\theta_{12},\delta) {\sf U}(t_N,t_D,\vec{p})]_{e2}|^2$ from $|U_{e2}(\theta_{23},\theta_{13},\theta_{12},\delta)|^2$ in terms of a regeneration factor,
\begin{align}
\label{eq:freg1}
|[U(\theta_{23},\theta_{13},\theta_{12},\delta) {\sf U}(t_N,t_D,\vec{p})]_{e2}|^2 \equiv |U_{e2}(\theta_{23},\theta_{13},\theta_{12},\delta)|^2+ F_{reg}\,.
\end{align}
As follows from equation \eqref{eq:evolopadiabatic}, for a Hamiltonian of the form \eqref{eq:Happox} the $(1,3)$ and $(3,1)$ as well as  $(2,3)$ and $(3,2)$ elements of ${\sf U}(t_N,t_D,\vec{p})$ are zero.  ${\sf U}_{33}(t_N,t_D,\vec{p}) = \exp[-i\phi(t_N,t_D,\vec{p})]$ further implies $|{\sf U}_{33}(t_N,t_D,\vec{p})|=1$. From unitarity of $U(\theta_{23},\theta_{13},\theta_{12},\delta)$ and ${\sf U}(t_N,t_D,\vec{p})$ it then follows that 
\begin{align}
\label{eq:freg2}
|[U(\theta_{23},\theta_{13},\theta_{12},\delta) {\sf U}(t_N,t_D,\vec{p})]_{e1}|^2  = |U_{e1}(\theta_{23},\theta_{13},\theta_{12},\delta)|^2 - F_{reg}\,.
\end{align}
Combining equations \eqref{eq:freg1} and \eqref{eq:freg2} we obtain
\begin{align}
|{\sf U}_{ee}(t_N,t_P,\vec{p})|^2 & = |{\sf U}_{ee}(t_D,t_P,\vec{p})|^2 \nonumber \\
& - (|U_{e1}(\theta_{23},\theta_{13},\theta^P_{12},\delta)|^2 - |U_{e2}(\theta_{23},\theta_{13},\theta^P_{12},\delta)|^2)F_{reg}\,.
\end{align}
In the standard parameterization of the PMNS matrix, see equation \eqref{eq:UPMNS}, 
\begin{align}
|{\sf U}_{ee}(t_N,t_P,\vec{p})|^2 & = |{\sf U}_{ee}(t_D,t_P,\vec{p})|^2 - c^2_{13}\cos(2\theta^P_{12})F_{reg}\,.
\end{align}
Using unitarity of ${\sf U}(t_N,t_D,\vec{p})$ and $U(\theta_{23},\theta_{13},\theta_{12},\delta)$, as well as taking into account that the $(3,2)$ element of ${\sf U}(t_N,t_D,\vec{p})$ is zero in the used approximation, we can cast the regeneration factor into the form 
\begin{align}
F_{reg} = c^2_{13}\,\mathrm{Re}\bigl({\sf U}_{12}(t_N,t_D,\vec{p})\bigl[\cos(2\theta_{12}){\sf U}^*_{12}(t_N,t_D,\vec{p}) + \sin(2\theta_{12}){\sf U}^*_{22}(t_N,t_D,\vec{p})\bigr]\bigr)\,.
\end{align}
The PREM model distinguishes nine layers with the matter density slowly varying inside the layers  and sharp density changes at the borders. The detailed density profile can be found in table II of reference \citep{DZIEWONSKI1981297}. Neutrino propagation within the layers is close to adiabatic, whereas at the borders the adiabaticity is strongly violated \citep{deHolanda:2004fd,Akhmedov:2004rq,Ioannisian:2017dkx}. Hence, for a neutrino crossing only the first layer
\begin{align}
\mathsf{U}(t_N,t_D,\vec{p}) = O_{12}(\vartheta_{12}(t_{N_-},\vec{p})) \exp[-i\phi(t_N,t_D,\vec{p})] O^\dagger_{12}(\vartheta_{12}(t_{D_+},\vec{p}))\,,
\end{align}
where $t_{D_+} = t_{D} + \epsilon$ is the moment shortly after the neutrino enters the Earth, and  $t_{N_-} = t_N-\epsilon$ is the moment shortly before it leaves it. To a good approximation, the matter density profile of the Earth is spherically symmetric, hence 
$\vartheta_{12}(t_{N_-},\vec{p}) = \vartheta_{12}(t_{D_+},\vec{p})$. In this approximation
\begin{align}
\label{eq:Fregonelayer}
F_{reg} = c^2_{13}\sin(2\vartheta_{12}(t_{D_+},\vec{p}))\sin(2\theta_{12}(t_{D_+},\vec{p}))\sin^2(\Delta\phi_{21}(t_N,t_D,\vec{p})/2)\,,
\end{align}
where $\theta_{12}(t,\vec{p}) \equiv \theta_{12} + \vartheta_{12}(t,\vec{p})$ and $\Delta\phi_{ij} \equiv \phi_i-\phi_j$. For a neutrino crossing the first and the second layers 
\begin{align}
\mathsf{U}(t_N,t_D,\vec{p}) & = O_{12}(\vartheta_{12}(t_{N_-},\vec{p})) \exp[-i\phi(t_N,t_2,\vec{p})] O^\dagger_{12}(\vartheta_{12}(t_{2_+},\vec{p}))\nonumber\\
& \times O_{12}(\vartheta_{12}(t_{2_-},\vec{p})) \exp[-i\phi(t_2,t_1,\vec{p})] O^\dagger_{12}(\vartheta_{12}(t_{1_+},\vec{p}))\nonumber\\
& \times O_{12}(\vartheta_{12}(t_{1_-},\vec{p})) \exp[-i\phi(t_1,t_D,\vec{p})] O^\dagger_{12}(\vartheta_{12}(t_{D_+},\vec{p}))\,,
\end{align}
where $t_1$ and $t_2$ (with $t_D < t_1 < t_2 < t_N$) are the moments the neutrino enters and leaves the second layer respectively. The spherical symmetry of the Earth implies $\vartheta_{12}(t_{1_-},\vec{p}) = \vartheta_{12}(t_{2_+},\vec{p})$ and $\vartheta_{12}(t_{1_+},\vec{p}) = \vartheta_{12}(t_{2_-},\vec{p})$. This results in the regeneration factor of the form
\begin{align}
\label{eq:Fregthreelayers}
F_{reg} & = c^2_{13} \bigl(\sin(2\vartheta_{12}(t_{D_+},\vec{p}))[\,\sin(\phi(t_1,t_D,\vec{p}))\cos(\phi(t_2,t_1,\vec{p})/2) \nonumber\\
&+ \cos(2\Delta\vartheta_{12}(t_1,\vec{p})) \cos(\phi(t_1,t_D,\vec{p}))\sin(\phi(t_2,t_1,\vec{p})/2)]\nonumber\\
&+ \cos(2\vartheta_{12}(t_{D_+},\vec{p})) \sin(2\Delta\vartheta_{12}(t_1,\vec{p}))\sin(\phi(t_2,t_1,\vec{p})/2) \bigr) \nonumber\\
&\times  \bigl(\sin(2\theta_{12}(t_{D_+},\vec{p}))[\,\sin(\phi(t_1,t_D,\vec{p}))\cos(\phi(t_2,t_1,\vec{p})/2) \nonumber\\
&+ \cos(2\Delta\vartheta_{12}(t_1,\vec{p})) \cos(\phi(t_1,t_D,\vec{p}))\sin(\phi(t_2,t_1,\vec{p})/2)]\nonumber\\
&+ \cos(2\theta_{12}(t_{D_+},\vec{p})) \sin(2\Delta\vartheta_{12}(t_1,\vec{p}))\sin(\phi(t_2,t_1,\vec{p})/2) \bigr)\,,
\end{align}
where $\Delta\vartheta_{12}(t_1,\vec{p}) \equiv \vartheta_{12}(t_{1_+},\vec{p}) - \vartheta_{12}(t_{1_-},\vec{p})$ is the jump of the diagonalization angle induced by the jump of the matter density at the border between the two layers. Replacing in equation \eqref{eq:Fregthreelayers} $\Delta\vartheta_{12}(t_1,\vec{p})\rightarrow 0$, $\phi(t_2,t_1,\vec{p}) \rightarrow 0$, and $\phi(t_1,t_D,\vec{p})\rightarrow \phi(t_1,t_D,\vec{p})/2$ we recover equation \eqref{eq:Fregonelayer}, as expected. 

The largest density jumps, from zero to $\sim 3.4\,\mathrm{g/cm}^3$ and from $\sim 5.5\,\mathrm{g/cm}^3$ to $\sim 9.9\,\mathrm{g/cm}^3$, occur within the first three layers (radius from 6371 km to 6346 km) and at the border between the mantle and the outer core (radius 3480 km) respectively. The density jumps between the other layers are considerably smaller. For the trajectories crossing more than three layers, the contribution of the short span occupied by the first three layers can be neglected. Therefore, the Earth can be approximated by a sphere of radius 6346 km with the density $\sim 3.4\,\mathrm{g/cm}^3$ at the `surface' and a density jump from $\sim 5.5\,\mathrm{g/cm}^3$ to $\sim 9.9\,\mathrm{g/cm}^3$ at the	 radius of 3480 km. In this approximation equation \eqref{eq:Fregonelayer} can be used for all trajectories crossing the mantle, while equation \eqref{eq:Fregthreelayers} can be used for all trajectories crossing also the core. Numerically,  this approximation yields acceptably accurate results, especially for the trajectories crossing many layers, see right panel of figure \ref{fig:daynightneutrinos}. 

\section{\label{sec:classicallimit}Classical limit}

In this appendix we provide an example a classical solution of the evolution equation for the Wigner function. For the neutrino propagating in a homogeneous medium the latter is given, in the ultra\-relativistic limit, by equation \eqref{eq:evoleqhompot},
\begin{align}
\label{eq:evoleqhompot_app}
(\partial_t+\vec{v}_\vec{p}\partial_\vec{x})\varrho(t,\vec{x},\vec{p}) = -i[{\sf H}(\vec{p}),\varrho(t,\vec{x},\vec{p})]\,.
\end{align}
It is solved by equation \eqref{eq:wignerfunctimedeppotappr},
\begin{align}
\label{eq:wignerfunctimedeppotappr_app}
\varrho(t,\vec{x},\vec{p}) = {\sf U}(t,t_P,\vec{p})  g(t,\vec{x},\vec{p}) {\sf U}^\dagger(t,t_P,\vec{p}) \,,
\end{align}
with the phase factor $ {\sf U}(t,t_P,\vec{p})$ given by equation \eqref{eq:evolutionoperator} and the shape factor by equation \eqref{eq:shapefactconstpotlowestord}. For Gaussian initial conditions the shape factor is given by equation \eqref{eq:shapewignergaussian},
\begin{align}
\label{eq:shapewignergaussian_app}
g_{\alpha\beta}(t,\vec{x},\vec{p}) = \delta_{\alpha e}\delta_{e\beta}\cdot 2^3\exp\biggl(-\frac{(\vec{p}-\vec{p}_w)^2}{2\sigma_p^2}\biggr)\,
\exp\biggl(-\frac{(\vec{v}_\vec{p}(t-t_P)-(\vec{x}-\vec{x}_P))^2}{2\sigma^2_x}\biggr)\,.
\end{align}
The momentum and coordinate uncertainties are related by $\sigma_x\sigma_p=\frac12$. However, equation \eqref{eq:wignerfunctimedeppotappr_app} solves equation \eqref{eq:evoleqhompot_app} also if we consider $\sigma_p$ and $\sigma_x$ as independent and, in particular, if we simultaneously take the limit $\sigma_p\rightarrow 0$ and $\sigma_x \rightarrow 0$. Using the relation 
\begin{align}
\delta(x)=\lim_{\epsilon\rightarrow 0}\frac{1}{\sqrt{2\pi\epsilon}}\exp\left(-\frac{x^2}{2\epsilon}\right)
\end{align}
we obtain for the shape factor in this limit 
\begin{align}
\label{eq:shapeclassical}
g_{\alpha\beta}(t,\vec{x},\vec{p}) \propto \delta_{\alpha e}\delta_{e\beta}\cdot\delta(\vec{p}-\vec{p}_w)\delta(\vec{v}_\vec{p}(t-t_P)-(\vec{x}-\vec{x}_P))\,.
\end{align}
Whereas the resulting Wigner function solves the evolution equation \eqref{eq:evoleqhompot_app}, it describes a state with a definite coordinate and momentum, and hence contradicts the Heisenberg uncertainty principle. Let us emphasize that the example presented in this appendix has no quantum description unless $\hbar \rightarrow 0$.

\section{\label{sec:densitymatrix}Initial conditions given by density matrix}
In section \ref{sec:qmapproach} the derivation of the evolution equation for the Wigner function was based on the as\-sump\-tion, that the neutrino is described by a pure state. Here we generalize the analysis to neutrinos described by a density matrix, 
\begin{align}
\hat{\mathscr{P}} = \sum_n p_n |\psi_{(n)}\rangle \langle \psi_{(n)}|\,,
\end{align}
where the non-negative coefficients $p_n$ add up to unity. Expressed in terms of the density matrix the Wigner function takes the form 
\begin{align}
\label{eq:Wignfuncdensmatr}
\varrho_{ij}(t,\vec{x},\vec{p}) = \int \frac{d^3\vec{\Delta}}{(2\pi)^3} e^{i\vec{\Delta}\vec{x}}\,
\mathrm{tr} [|\vec{p}-\vec{\Delta}/2\rangle \langle \vec{p}+\vec{\Delta}/2| \mathscr{P}(t)]\,.
\end{align}
Its time derivative is proportional to time derivative of the density matrix. The latter is determined by the Liouville -- von Neumann equation,
\begin{align}
\label{eq:LvNeq}
\partial_t\hat{\mathscr{P}}(t) = -i [\hat{\mathscr{H}}(t),\hat{\mathscr{P}}(t)]\,.
\end{align}
Combining equations \eqref{eq:Wignfuncdensmatr} and \eqref{eq:LvNeq} we arrive at 
\begin{align}
\label{eq:eveq1}
\partial_t\varrho_{ij}(t,\vec{x},\vec{p}) = -i \int \frac{d^3\vec{\Delta}}{(2\pi)^3} e^{i\vec{\Delta}\vec{x}}\,
\mathrm{tr} [|\vec{p}-\vec{\Delta}/2\rangle \langle \vec{p}+\vec{\Delta}/2| [\hat{\mathscr{H}}(t),\hat{\mathscr{P}}(t)]]\,.
\end{align}
The Hamiltonian operator is the sum of the kinetic and potential energy operators, $\hat{\mathscr{H}} = \hat{\mathscr{K}} + \hat{\mathscr{V}}$. The former is diagonal in momentum space, $\langle \vec{p} |\hat{\mathscr{K}}| \vec{p}' \rangle = (2\pi)^3\delta(\vec{p}-\vec{p}')\mathsf{K}(\vec{p})$, where $\mathsf{K}$ is diagonal in  the generation space. Hence its contribution to the right-hand side of equation \eqref{eq:eveq1} is given by
\begin{align}
\label{eq:eveq2}
\partial_t\varrho_{ij}(t,\vec{x},\vec{p}) & \ni -i \int \frac{d^3\vec{\Delta}}{(2\pi)^3} e^{i\vec{\Delta}\vec{x}}\,
\mathrm{tr} [|\vec{p}-\vec{\Delta}/2\rangle \langle \vec{p}+\vec{\Delta}/2| \nonumber \\
& \times [
\mathsf{K}_{ii}(\vec{p}+\vec{\Delta}/2) \hat{\mathscr{P}}_{ij}(t) -  \hat{\mathscr{P}}_{ij}(t) \mathsf{K}_{jj}(\vec{p}-\vec{\Delta}/2)
]]\,.
\end{align}
The potential energy operator is diagonal in coordinate space, $\langle \vec{x} |\hat{\mathscr{V}}(t)| \vec{x}' \rangle = \delta(\vec{x}-\vec{x}')\mathsf{V}(t,\vec{x})$. To compute its contribution it is convenient to express $|\vec{p}\mp\vec{\Delta}/2\rangle$ in equations \eqref{eq:Wignfuncdensmatr} and \eqref{eq:eveq1} in terms of their coordinate space counterparts. Using $|\vec{p}\rangle = \int d^3\vec{x}\, e^{i\vec{p}\vec{x}}\, |\vec{x}\rangle$ we obtain
\begin{align}
\label{eq:Wignfuncdensmatr1}
\varrho_{ij}(t,\vec{x},\vec{p}) =  \int d^3\vec{y}e^{i\vec{p}\vec{y}}\,
\mathrm{tr} [|\vec{x}-\vec{y}/2\rangle \langle \vec{x}+\vec{y}/2| \mathscr{P}(t)]\,
\end{align}
and 
\begin{align}
\partial_t\varrho_{ij}(t,\vec{x},\vec{p}) = -i \int d^3\vec{y}e^{i\vec{p}\vec{y}}\,
\mathrm{tr} [|\vec{x}-\vec{y}/2\rangle \langle \vec{x}+\vec{y}/2| [\hat{\mathscr{H}}(t),\hat{\mathscr{P}}(t)]]\,
\end{align}
respectively. The contribution of the potential energy operator is given by
\begin{align}
\label{eq:eveq3}
\partial_t\varrho_{ij}(t,\vec{x},\vec{p}) & \ni -i \int d^3\vec{y}e^{i\vec{p}\vec{y}}\,
\mathrm{tr} [|\vec{x}-\vec{y}/2\rangle \langle \vec{x}+\vec{y}/2| \nonumber \\
& \times [
\mathsf{V}_{ik}(t,\vec{x}+\vec{y}/2) \hat{\mathscr{P}}_{kj}(t) -  \hat{\mathscr{P}}_{ik}(t) \mathsf{V}_{kj}(t,\vec{x}-\vec{y}/2)
]]\,.
\end{align}
Expanding the kinetic and potential terms, $\mathsf{K}(\vec{p}\pm\vec{\Delta}/2) \approx \mathsf{K}(\vec{p})\pm \frac12\vec{\Delta} \partial_\vec{p}\mathsf{K}(\vec{p})$ and $\mathsf{V}(t,\vec{x}\pm\vec{y}/2) \approx \mathsf{V}(t,\vec{x}) \pm \frac12 \vec{y} \partial_\vec{x}\mathsf{V}(t,\vec{x})$, using equations \eqref{eq:Wignfuncdensmatr} and \eqref{eq:Wignfuncdensmatr1}, and combining equations \eqref{eq:eveq2} and \eqref{eq:eveq3} we recover equation \eqref{eq:timederrhoingom3}.

%\bibliographystyle{JHEP}
%\bibliography{references}

\providecommand{\href}[2]{#2}\begingroup\raggedright\endgroup

\end{document}